\documentclass[]{aa}  

\usepackage{natbib}
\bibpunct{(}{)}{;}{a}{}{,}
\usepackage{graphicx}
\usepackage{revsymb}	
\usepackage{amsmath}	
\usepackage[usenames]{color}	
\usepackage{epstopdf}	
\usepackage{txfonts}
\usepackage{amssymb}
\usepackage{color}
\usepackage[justification=centering]{caption}
\usepackage{geometry} 
\usepackage[parfill]{parskip} 
\usepackage{amssymb}
\usepackage{slantsc}
\usepackage{array}
\usepackage{url}
\usepackage{amsfonts}
\usepackage[colorlinks=true,linkcolor=black, urlcolor=black, citecolor=black]{hyperref}
\usepackage{hyperref} 
\usepackage{sidecap}
\usepackage{graphicx}
\usepackage{xcolor}
\usepackage{nicefrac}
\usepackage{subfigure}
\usepackage{epsfig}
\colorlet{rouge}{red!70!darkgray}

\begin{document}
\title{Combining multiple structural inversions to constrain the Solar modelling problem}
\author{G. Buldgen\inst{1}\and S. J. A. J. Salmon\inst{2}\and A. Noels\inst{2} \and R. Scuflaire\inst{2} \and J. Montalban\inst{3} \and V.A. Baturin\inst{4} \and  P. Eggenberger\inst{5} \and V.K. Gryaznov\inst{6,7} \and I.L. Iosilevskiy\inst{8,9} \and G. Meynet\inst{5} \and W. J. Chaplin\inst{1} \and A. Miglio\inst{1} \and A.V. Oreshina\inst{4} \and O. Richard\inst{10} \and A.N. Starostin\inst{11}}
\institute{School of Physics and Astronomy, University of Birmingham, Edgbaston, Birmingham B15 2TT, UK. \and STAR Institute, Université de Liège, Allée du Six Août 19C, B$-$4000 Liège, Belgium \and Dipartimento di Fisica e Astronomia, Università degli Studi di Padova, vicolo dell’Osservatorio 3, 35122 Padova, Italy \and Sternberg Astronomical Institute, Lomonosov Moscow State University, 119234 Moscow, Russia \and Observatoire de Genève, Université de Genève, 51 Ch. Des Maillettes, CH$-$1290 Sauverny, Suisse \and Institute of Problems of Chemical Physics RAS, 142432 Chernogolovka, Russia \and Tomsk State University, 634050 Tomsk, Russia \and Joint Institute for High Temperatures RAS, 125412 Moscow, Russia \and Moscow Institute of Physics and Technology, 141701 Dolgoprudnyi, Russia\and LUPM, Université de Montpellier, CNRS, Place E. Bataillon, 34095 Montpellier Cedex, France \and Troitsk Institute for Innovation and Fusion Research, 142190 Troitsk, Russia}
\date{March, 2018}
\abstract{The Sun is the most studied of all stars, which serves as a reference for all other observed stars in the Universe. Furthermore, it also serves the role of a privileged laboratory of fundamental physics and can help us better understand processes occuring in conditions irreproducible on Earth. However, our understanding of our star is currently stained by the so-called solar modelling problem, resulting from comparisons of theoretical solar models to helioseismic constraints. These discrepancies can stem from various causes, such as the radiative opacities, the equation of state as well as the mixing of the chemical elements.}
{By analysing the potential of combining information from multiple seismic inversions, we aim to help disentangling the origins of the solar modelling problem.}
{We combine inversions of the adiabatic sound speed, an entropy proxy and the Ledoux discriminant with other constraints such as the position of the base of the convective zone and the photospheric helium abundance. First, we test various combinations of standard ingredients available for solar modelling such as abundance tables, equation of state, formalism for convection and diffusion and opacity tables. Second, we study the diagnostic potential of the inversions on models including ad-hoc modifications of the opacity profile and additional mixing below the convective envelope.}{We show that combining inversions provides stringent constraints on the required modifications to the solar ingredients, far beyond what is achievable only from sound speed inversions. We constrain the form and amplitude of the opacity increase required in solar models and show that a 15$\%$ increase at $\log T =6.35$ provides a significant improvement but is on his own insufficient. A more global increase of the opacity, within the uncertainties of the current tables, coupled with a localized additional mixing at the bottom of the convective zone provides the best agreement for low metallicity models. We show that high metallicity models do not satisfy all the inversion results. We conclude that the solar modelling problem likely occurs from multiple small contributors, as other ingredients such as the equation of state or the formalism of convection can induce small but significant changes in the models and that using phase shift analyses combined with our approach is the next step for a better understanding of the inaccuracies of solar models just below the convective envelope.}{}
\maketitle
\section{Introduction}

In the past decades, helioseismology has been one of the most successful fields of astrophysics. Its achievements stem from the excellent quality of the seismic data, allowing for thorough comparisons between solar models and the Sun. These studies led to the precise determination of the position of the base of the solar convective zone \citep{KosovBCZ,JCD91Conv,Basu97BCZ} (hereafter BCZ), the present day helium abundance in the convective zone \citep{Vorontsov91,Dziembowski91,BasuYSun, RichardY,Vorontsov13}, unconstrained by photosphere spectroscopy, the determination of the radial profile of thermodynamical quantities inside the Sun \citep[see e.g.][and references therein]{Antia94, Marchenkov} and that of the solar rotation profile \citep{BrownRota, KosovichevRota, SchouRota, Garcia2010}. 

These successes paved the way for the asteroseismic analyses of solar-like oscillators and indirectly established the study of stellar oscillations as the ``golden path'' to determine stellar fundamental parameters such as mass, radius and age. However, while the reliability of seismology as a test of solar and stellar structure is well established, the accuracy of stellar and solar models is still under question. In the solar case, the most tedious issue is linked to the revision of the abundance of heavy elements by \citet{AGS04O,AGS05C,AGSS09,Caffau,Scott2015I,Scott2015II,Grevesse2015}, which leads to a strong disagreement between solar models and helioseismology which, in turn, leads to question the reliability of solar models. The main difficulty with this issue is that the observed discrepancies can originate from multiple sources. Amongst them, the alleged largest contributor is the radiative opacity, which is thought to be underestimated for the physical conditions of the BCZ \citep{Bailey,Iglesias2015,Nahar,Pradhan,Zhao,Pain2018}, although the origin of the discrepancy is still under debate \citep{Blancard2016,Iglesias2017,Pain2017}. However, other inputs of standard models can also contribute to the discrepancies, such as the equation of state, the formalism used for microscopic diffusion, the formalism of convection or the nuclear reaction rates. Hence, the recipe of the standard model itself is a limiting factor to its expected accuracy. For example, the modelling of the BCZ, where the interplay between various so-called ``non-standard'' processes, such as rotation, turbulence and magnetism leads to the apparition of the so-called the solar tachocline\footnote{The tachocline is defined, by analogy with the thermocline in oceanography, as the transition region in the Sun from the differentially latitudinal rotating convective envelope to the rigidly rotating radiative region \citep{SpiegelZahn1992}.}. The current inability to accurately model these processes implies that the reliability of the standard solar models in these regions is quite questionable.  
	  
In previous papers \citep{BuldgenA,BuldgenS}, we presented new structural inversions and shown how these could be used to re-analyse the solar modelling problem. We show how combining all these inversions together can be used to offer stringent constraints on the processes and changes that can be applied to standard models to solve their current issue with seismic constraints. This is done by disentangling the competing effects of the various uncertainties on the physical processes inside the Sun. A similar analysis can be found in \citet{Ayukov2011} and \citet{Ayukov2017}, where in this last study, the problem is redefined as an expanded solar calibration procedure.

We start in Sect. \ref{SecCombinedInversions} by briefly presenting the ingredients of the current standard solar models with their respective uncertainties and describe in Sect. \ref{SecInversionsSTD} the set of models we have computed using the Liège stellar evolution code (CLES) with various physical ingredients. We then present the adiabatic sound speed, entropy proxy and Ledoux discriminant inversions for our standard solar models and discuss the observed variations as well as other potential improvements that could lead to significant changes at the level of accuracy of helioseismic investigations. In Sect. \ref{SecNonStdModels}, we compute solar models including ad-hoc modifications, using parametric variations in the opacity profile and extra-mixing of the chemical elements below the convective zone to test the potential impact of such processes in the reconciliation of solar models with helioseismic constraints. 

\section{Combined inversions for standard solar models}\label{SecCombinedInversions}

In this section, we present structural inversion results of various thermodynamical quantities and discuss how their combination on a sample of standard solar models provides an in-depth analysis of the solar modelling problem. 

The use of structural inversions in global helioseismology is now a standard approach. In this work, we use the linear formulation of the inverse problem following the developments of \citet{Dziemboswki90}, based on the variational analysis of the pulsation equations \citep[see e.g.][and references therein]{LyndenBell}. These developments lead to an inverse problem of the form
\begin{align}
\frac{\delta \nu_{n,\ell}}{\nu_{n,\ell}}=\int_{0}^{R}K^{n,\ell}_{s_{1},s_{2}}\frac{\delta s_{1}}{s_{1}}dr + \int_{0}^{R}K^{n,\ell}_{s_{2},s_{1}}\frac{\delta s_{2}}{s_{2}}dr + \mathcal{F}_{\mathrm{Surf}}, \label{EqInversion}
\end{align}
with $\frac{\delta \nu_{n,\ell}}{\nu_{n,\ell}}$ the relative frequency differences of degree $\ell$ and radial order $n$,$ \frac{\delta s_{i}}{s_{j}}$ the relative differences in the acoustic variables for the considered formulation of the inverse problem, $K^{n,\ell}_{s_{i},s_{j}}$ the kernel functions associated with the acoustic variables and $\mathcal{F}_{\mathrm{Surf}}$ an operator describing the surface regions, unaccurately described with the hypotheses of the variational approach. We consider a description of these ``surface effects'' as a $6^{\mathrm{th}}$ order polynomial as in \citet{RabelloParam}. We use the SOLA inversion method from \citet{Pijpers} to solve Eq. \ref{EqInversion}.

We use three pairs of acoustic variables for which the integral relations are solved. First, we carry out inversions of the squared adiabatic sound-speed, denoted $c^{2}$, using the $(c^{2},\rho)$ structural kernels. Second, we carry out inversions for an entropy proxy \citep{BuldgenS}, denoted $S_{5/3}=\frac{P}{\rho^{5/3}}$, which stems from the Sackur-Tetrode equation for the entropy of a mono-atomic non-degenerate ideal gas. The advantage of this quantity is that it reproduces the plateau-like behaviour expected from the entropy in fully mixed regions with a nearly-adiabatic stratification such as the deep convective layers in the Sun. The height of this plateau is a key parameter of the solar convective zone. The height of the entropy proxy plateau is very sensitive to the temperature gradient in the radiative zone, which can thus be strongly constrained. For this inversion, we used the $(S_{5/3},\Gamma_{1})$ structural kernels. Finally, we also invert for the Ledoux discriminant, denoted $A=\frac{d \ln P}{d \ln r}-\frac{1}{\Gamma_{1}}\frac{d \ln \rho}{d \ln r}$. The Ledoux discriminant has the advantage of being much more sensitive to local variations. It allows to analyse the properties of both the temperature and chemical composition gradient at the BCZ, offering strong constraints on potential extra-mixing processes.

Inversions were performed individually using the calibrated model as a reference to compute the structural kernels and individual frequencies. In Sect. \ref{SecModels}, we describe how these quantities help us analyse the impact of various ingredients of standard solar models. The inversion results are presented for the whole set of models of Sect. \ref{SecModels} and we also present in table \ref{tabSTDModels} other key helioseismic constraints such as the BCZ and the mass coordinate at this position as well as the helium and heavy elements abundances in the convective zone and the initial values of these quantities for each of our models.

The solar models were computed using the Liège stellar evolution code \citep[CLES,][]{ScuflaireCles} and their eigenfrequencies were computed using the Liège oscillation code \citep[LOSC,][]{ScuflaireOsc}. The seismic data used for the inversions is a combination of BiSON and MDI data \citep[see][]{BasuSun, Davies}. Inversions were carried out using an adapted version of the InversionKit software \citep{ReeseDens}.

\subsection{Standard solar models and their physical ingredients}\label{SecModels}

The definition of the standard solar model stems from \citet{Bahcall82} and defines a well-posed mathematical problem to compute a theoretical model of the Sun. A standard solar model is a $1\mathrm{M}_{\odot}$ stellar model, evolved up to the solar age, reproducing the current photospheric ratio of heavy elements over hydrogen, $Z/X$, the solar radius and the solar effective temperature (or luminosity). To fulfill these constraints, th models are built using three free parameters, the solar initial hydrogen and heavy elements abundance and the mixing length parameter of convection. For the calibrations considered in this paper, we used the solar parameters of \citet{Mamajek}.

Besides the mathematical setup of the problem, standard models are also defined by a set of physical ingredients, like the metal mixture composition of the stellar plasma, the equation of state, the radiative opacities and the nuclear reaction rates. The only transport processes included in standard solar models are thermal convection and microscopic diffusion, often using rather simple approaches. Convection follows local and simplified formalisms, such as the standard mixing length theory (MLT) \citep{Cox} or the Full Spectrum of Turbulence model (FST) of \citet{Canuto91, Canuto, Canuto96}, while diffusion often uses approximations such as in \citet{Proffitt} and \citet{Thoul}.

While such a representation gives a satisfactory agreement with the Sun, it is still uncertain. The modelling of the interaction between various physical processes such as rotation, magnetism and turbulence at the BCZ, the radiative opacities and our depiction of convection in the upper layers of the solar envelope can be listed as the most uncertain aspects of the present state of solar modelling. However, other key ingredients of the internal solar structure such as the equation of state and the hypotheses we use to compute the diffusion of chemical elements and the cross-sections of some nuclear reactions which still present large uncertainties may also have a significant impact at the level of precision of helioseismic constraints \citep[see][and references therein for a discussion on some specific ingredients]{Boothroyd03}. In addition, improving the outer layers of the solar models can also have a slight effect on the solar modelling problem, as discussed in \citet{Gabriel96} and \citet{Schlattl97}. To that extent, the importance of the inclusion of turbulent pressure in the modelling of the outer convective layers and the traces it could leave in seismic inversions could be investigated \citep[see e.g.][]{Houdek2017,Sonoi2017}.

Moreover, the solar standard model neglects effects of physical processes such as rotation, the magnetic field, mass loss, internal gravity waves, compressible turbulent convection or overshooting\footnote{We denote as overshooting here the extent of the convective region beyond the formal Schwarzschild limit derived from the local convection theory used in the model.} and is unidimensional while some of the mentioned processes are intrinsically not. It is unable to reproduce the lithium depletion and the departures from spherical symmetry observed in the Sun. Some refinements to the transport of chemicals by diffusion are also often missing, such as the effects of radiative levitation or partial ionization. These effects have been studied for example by \citet{Turcotte} and \citet{Gorshkov10}. Other effects, such as the fineness of the opacity tables \citep[See e.g.][]{LePennec} or the choice of low-temperature opacity tables \citep{Guzik2010} will also impact the structure of solar models at a level seen by helioseismology. The impact of these so-called non-standard processes is expected to be small, but likely not negligible.

\subsection{Set of models and inversion results}\label{SecInversionsSTD}

To compare our standard models, we chose to keep one set of ingredients as reference and plot all other inverted profile in figures including this specific reference, to see directly the effects of various ingredients. We used a model built using the AGSS$09$ abundances \citep{AGSS09}, the FreeEOS equation of state \citep{Irwin}, the opal opacities \citep{OPAL}, the mixing-length theory of convection \citep{Cox}, the formalism for microscopic diffusion by \citet{Thoul} and the nuclear reaction rates from \citet{Adelberger}. All models also include effects of conduction from \citet{Potekhin} and from \citet{Cassisi} as well as low-temperature opacities from \citet{Ferguson}. We used grey atmosphere models in the Eddington approximation in all our models. 

\begin{figure*}
	\centering
		\includegraphics[width=17cm]{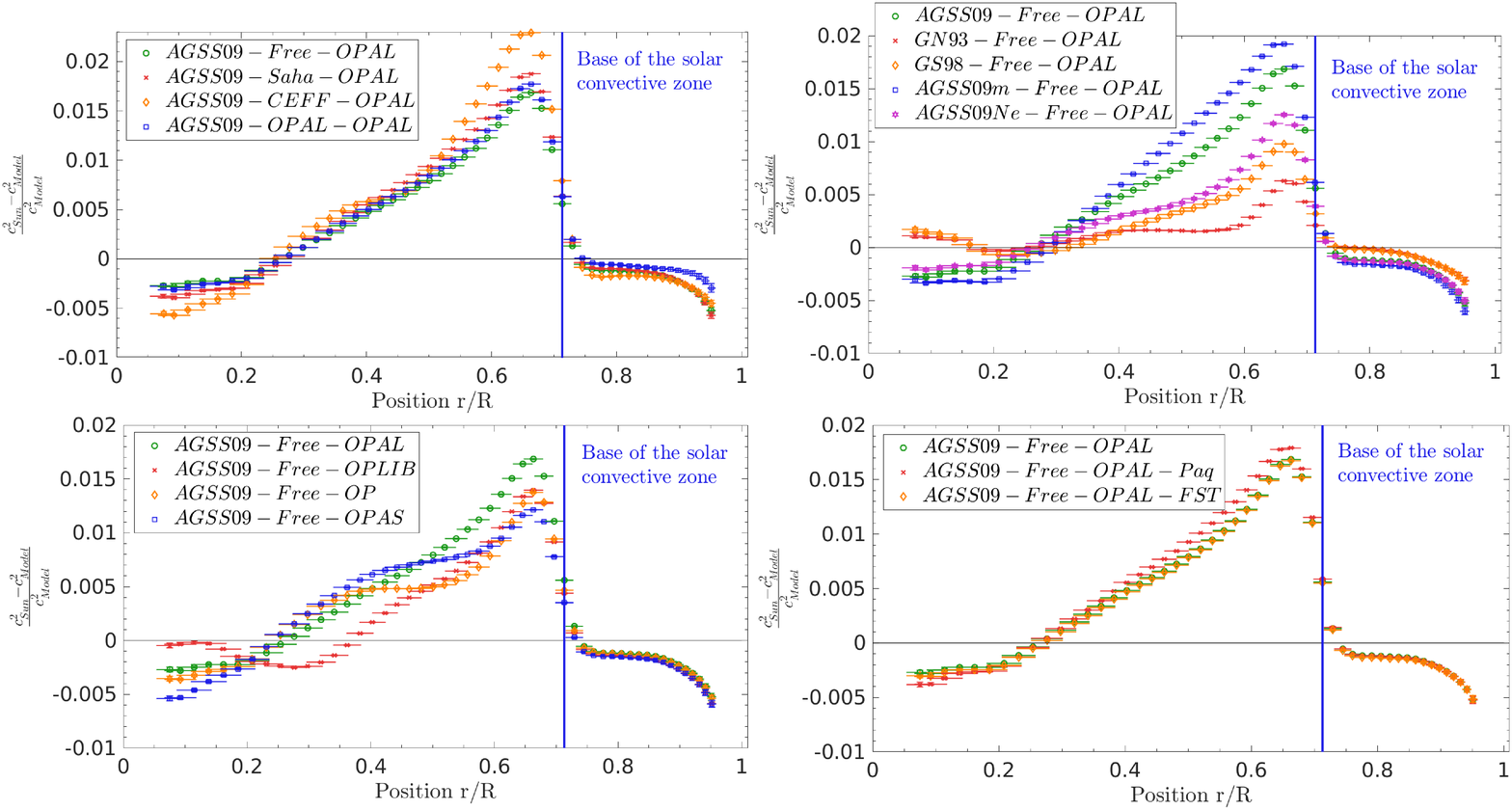}
	\caption{Relative differences in squared adiabatic sound speed between the Sun and calibrated solar models. Upper-left panel: effects of the equation of state. Upper-right panel: effects of abundance variations. Lower-left panel: effects of the opacity tables. Lower-right panel: effects of changing the diffusion coefficients and the treatment of convection.}
		\label{FigSoundSpeedSTD}
\end{figure*}

\begin{figure*}
	\centering
		\includegraphics[width=17cm]{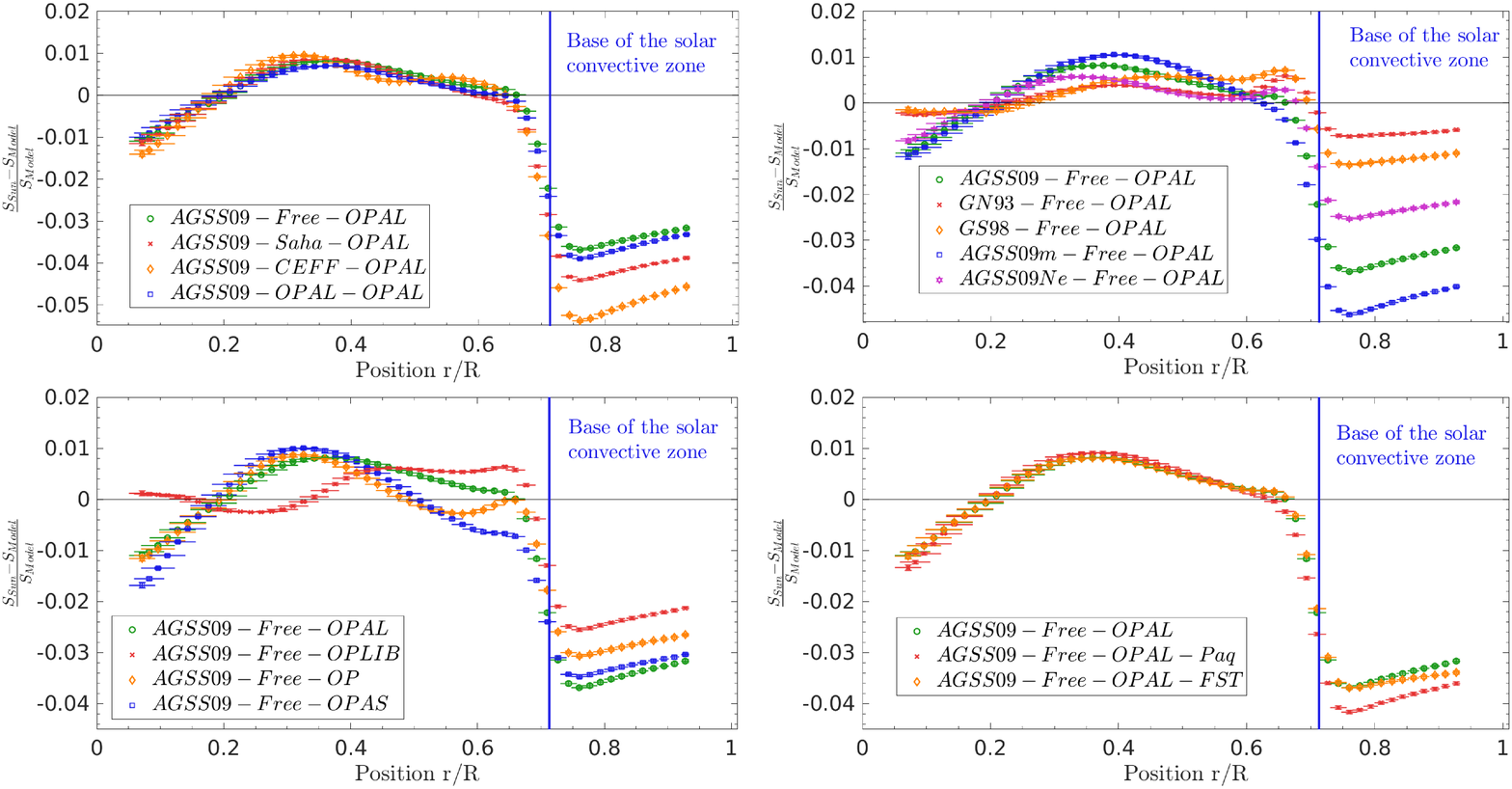}
	\caption{Relative differences in entropy proxy, $S_{5/3}$, between the Sun and calibrated solar models. Upper-left panel: effects of the equation of state. Upper-right panel: effects of abundance variations. Lower-left panel: effects of the opacity tables. Lower-right panel: effects of changing the diffusion coefficients and the treatment of convection.}
		\label{FigEntropyStd}
\end{figure*}

\begin{figure*}
	\centering
		\includegraphics[width=17cm]{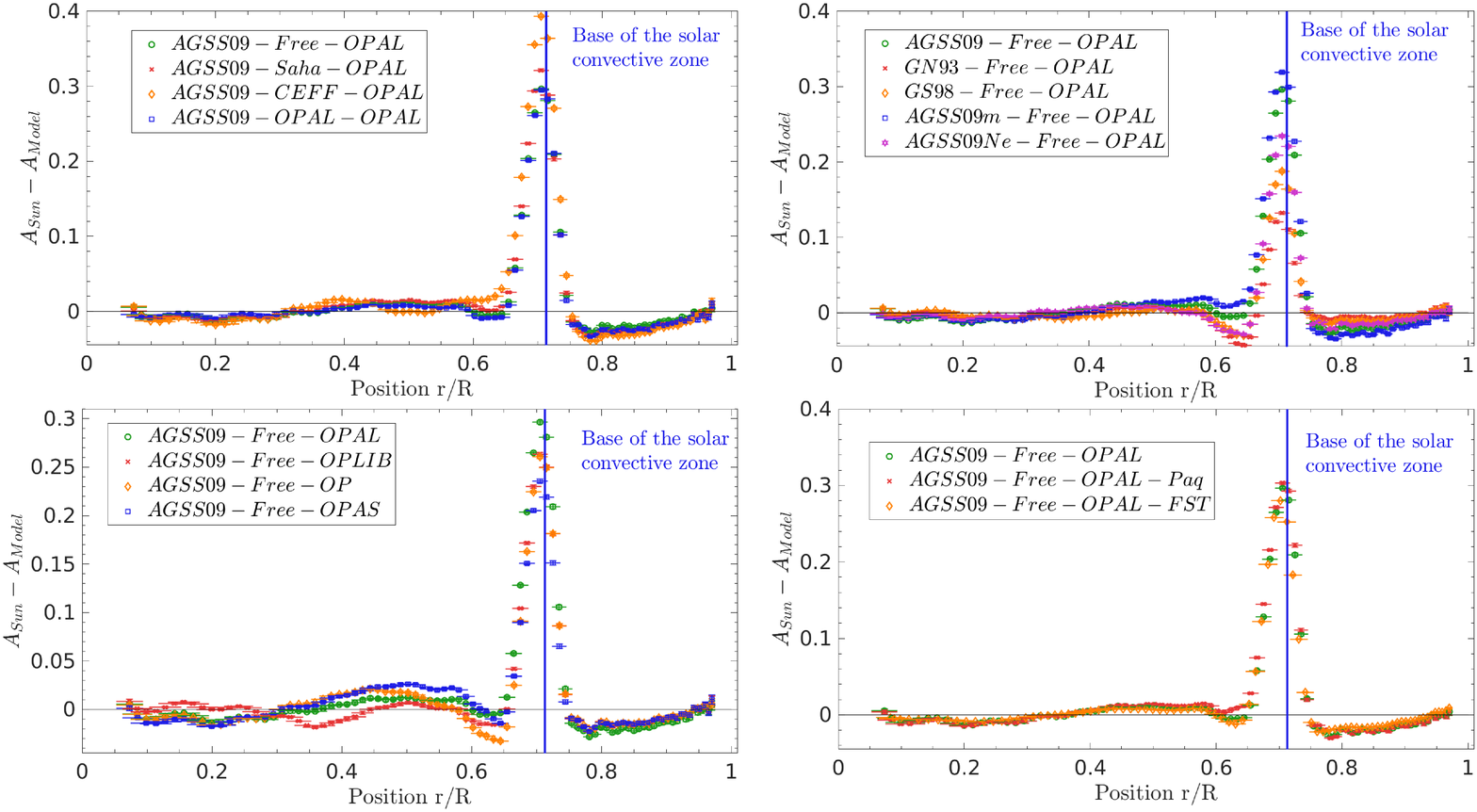}
	\caption{Relative differences in Ledoux discriminant, $A$, between the Sun and calibrated solar models. Upper-left panel: effects of the equation of state. Upper-right panel: effects of abundance variations. Lower-left panel: effects of the opacity tables. Lower-right panel: effects of changing the diffusion coefficients and the treatment of convection.}
		\label{FigLedouxSTD}
\end{figure*}

We subdivided our comparisons into four main effects: changing the equation of state, changing the opacity tables, changing the abundances, and changing the formalism for convection and diffusion. Each of these effects is respectively represented as a subpanel in Figs. \ref{FigSoundSpeedSTD}, \ref{FigEntropyStd} and \ref{FigLedouxSTD} illustrating the inversions of the squared adiabatic sound speed, the entropy proxy and the Ledoux convective parameters for each model of our sample. We use the SAHA-S \citep{Gryaznov06, Baturin,Gryaznov13}, the OPAL \citep{Rogerseos}, the FreeEOS and the CEFF equations of state \citep{CEFF, Irwin} to analyze the variations induced by changing the equation of state in the solar models.The EOS-tables used in the model computations are initially defined with the hydrogen $X$ and the total heavy elements mass fraction $Z$ as parameters of the chemical composition. To test ``standard'' opacity tables, we used models built with the OPAL, OPAS \citep{Mondet}, OPLIB \citep{Colgan} and OP \citep{Badnell} opacity tables. As for the abundances, we used models built with the former GN$93$ and GS$98$ abundances \citep{GrevNoels,GreSauv} and models built with the more recent AGSS$09$ abundances. We also computed one table for which the abundances of C, N, O, Ne and Ar were changed to the meteoritic ones as done in \citet{SerenelliComp}, denoted AGSS$09$m and one for which the recently suggested $40\%$ Neon abundance increase was taken into account \citep{Landi,Young}, denoted AGSS$09$Ne. For each of these composition tables, the solar $Z/X$ ratio to be reproduced was adapted accordingly and opacity tables were recomputed for each abundance table. Finally, we also considered using the diffusion coefficients from \citet{Paquette} instead of those from \citet{Thoul} and the FST formulation of \citet{Canuto96} besides models computed using the classical mixing-length theory \citep{Cox}. 

The physical ingredients of the models are summarized in table \ref{tabSTDModels} alongside the photospheric helium and heavy-elements abundances, the position of their BCZ, the mass coordinate at this position and their initial heavy-elements and helium abundances. The position of the BCZ can be directly compared to the helioseismic value of $0.713$ \citep{Basu97BCZ}. As for the photospheric helium abundance, we consider that a value above $0.245$ is acceptable as it agrees with intervals found in most recent studies \citep{Vorontsov13, VorontsovSolarEnv2014} and the usual value of $0.2485$ \citep{Antia94,Basu04}. The parameters of the reference AGSS$09$-FreeEOS-OPAL model mentioned earlier are given in the first line of table \ref{tabSTDModels} and the inversion results of this model are plotted in green throughout the paper and will be referred in the plots as AGSS$09$-Free-Opal. 

\begin{table*}[t]
\caption{Parameters of the standard solar models used in this study}
\label{tabSTDModels}
  \centering
  \resizebox{\linewidth}{!}{%
\begin{tabular}{r | c | c | c | c | c | c | c | c | c | c}
\hline \hline
\textbf{$\left(r/R\right)_{BCZ}$}&\textbf{$\left( m/M \right)_{CZ}$} &\textbf{$Y_{CZ}$}&\textbf{$Z_{CZ}$}&\textbf{$Y_{0}$}&\textbf{$Z_{0}$}&\textbf{EOS}&\textbf{Opacity}&\textbf{Abundances} & \textbf{Diffusion} & \textbf{Convection}\\ \hline
$0.7224$&$0.9785$&$0.2363$&$0.01361$&$0.2664$& $0.01511$ & FreeEOS & OPAL & AGSS09 & Thoul & MLT\\
$0.7230$&$0.9786$&$0.2376$&$0.01355$&$0.2685$& $0.01523$ & OPAL & OPAL & AGSS09 & Thoul & MLT\\
$0.7272$&$0.9799$&$0.2368$&$0.01360$&$0.2682$& $0.01515$& CEFF & OPAL & AGSS09 & Thoul & MLT\\
$0.7239$&$0.9790$&$0.2380$&$0.01355$&$0.2690$& $0.01524$ & SAHA-S & OPAL & AGSS09 & Thoul & MLT\\
$0.7215$&$0.9781$&$0.2350$&$0.01363$&$0.2647$& $0.01511$ & FreeEOS & OP & AGSS09 & Thoul & MLT\\ 
$0.7205$&$0.9777$&$0.2300$&$0.01372$&$0.2588$& $0.01520$ & FreeEOS & OPLIB & AGSS09 & Thoul & MLT\\ 
$0.7196$&$0.9779$&$0.2322$&$0.01368$&$0.2614$& $0.01516$ & FreeEOS & OPAS & AGSS09 & Thoul & MLT\\
$0.7224$&$0.9785$&$0.2363$&$0.01361$&$0.2664$& $0.01511$ & FreeEOS & OPAL & AGSS09 & Thoul & FST\\
$0.7235$&$0.9788$&$0.2373$&$0.01359$&$0.2648$& $0.01480$ & FreeEOS & OPAL & AGSS09 & Paquette & MLT\\
$0.7131$&$0.9757$&$0.2453$&$0.01809$&$0.2750$& $0.01999$ & FreeEOS & OPAL & GN93 & Thoul & MLT\\
$0.7157$&$0.9764$&$0.2465$&$0.01706$&$0.2765$& $0.01887$ & FreeEOS & OPAL & GS98 & Thoul & MLT\\
$0.7248$&$0.9789$&$0.2338$&$0.01343$&$0.2639$& $0.01496$ & FreeEOS & OPAL & AGSS09m & Thoul & MLT\\
$0.7207$&$0.9780$&$0.2373$&$0.01393$&$0.2655$& $0.01547$ & FreeEOS & OPAL & AGSS09Ne & Thoul & MLT\\
\hline
\end{tabular}
}
\end{table*}

The Inversions have been computed using the SOLA method \citep{Pijpers} and the guidelines of \citet{RabelloParam} using the softwares and equations of \citet{BuldgenA,BuldgenS}. The results are shown in Fig. \ref{FigSoundSpeedSTD} for the squared adiabatic sound speed, in Fig. \ref{FigEntropyStd} for the entropy proxy, $S_{5/3}$ and in Fig. \ref{FigLedouxSTD} for the Ledoux discriminant, $A$.

\subsection{Discussion}

From a first glance at the inversion results presented in Figs \ref{FigSoundSpeedSTD}, \ref{FigEntropyStd} and \ref{FigLedouxSTD}, the first unsurprising conclusion that can be drawn is that there is, to this day, no combination of opacity tables and equation of state that can reconciliate standard models computed using the AGSS$09$ abundances with helioseismic results. Moreover, one can note that using more recent opacity tables like the OP, OPAS or OPLIB tables does not lead to an unequivocal improvement of the solar modelling problem. 

\subsubsection{Changing the opacity tables}

We first study the effect of the opacity tables. The lower left panel of Figs \ref{FigSoundSpeedSTD}, \ref{FigEntropyStd} and \ref{FigLedouxSTD} show that the inversion results are usually better when using more recent opacity tables, especially for sound speed. However, as seen from the $Y_{S}$ values in table \ref{tabSTDModels}, this improvement is mitigated by the large decrease of the helium abundance in the convective zone. This effect is particularly strong for the OPLIB tables which give opacity values significantly lower than all other tables in the radiative zone. This implies an increase in the initial hydrogen abundance in the calibrated model to compensate for the opacity decrease, which in turn leads to a decrease in the helium abundance. This effect is also seen in the OP and OPAS models which all have a higher hydrogen abundance than the OPAL one. 

Some of these differences can be attributed to the various equations of state used in the computation of these opacity tables. Indeed, it will control the ionization stage of the elements and hence the radiative opacity. To get the ionization stage and ions distribution one needs of a basic thermodynamic model. It is commonly known that the OPAL opacities are based on  the ``physical picture'' approach \citep{Rogerseos} while the OP opacity tables used the so-called ``chemical picture'', and are essentially based on the MHD EOS code \citep{MHDI,MHDII,MHDIII,MHDIV}. Both the OPAS and OPLIB opacities used their own equation of state, also based on the chemical picture. It is interesting to note that both the OPAS and OPLIB tables find a slightly lower opacity in the lower radiative region, which disagrees with previous tables.

In addition, differences can also be observed in the number of metals considered. For example the OPAL opacities consider $21$ elements, the OP tables consider $17$ elements while the OPLIB and OPAS consider $30$ and $22$ elements respectively. 

\subsubsection{Changing the equation of state}

By testing various equations of state, it can be seen that the inversion results are significantly improved when using the OPAL, FreeEOS or SAHA-S equation of state, drawn in blue, green and red, respectively, when compared to the CEFF equation of state, in orange. However, differences in the chemical composition of the convective envelope can be seen between the models. In that sense, the SAHA-S and OPAL models seem to be very similar, with the only exception being the position and mass coordinate of the convective envelope. The FreeEOS and CEFF models, on the contrary, show larger discrepancies. In the case of FreeEOS, the differences are essentially found in the present helium abundance in the convective zone, while the BCZ is very similar. For CEFF, the differences are more striking and are seen in every parameter of the envelope and inverted profiles.

Intrinsic differences in EOS quantities may be related to the different ``first-principle'' approaches, as well as to the additional terms and refinements which enter the thermodynamic potential and the numerical techniques used to compute the equation of state. The OPAL equation of state and the OPAL opacity tables are based on the so-called ``physical picture''.  In this case, only fundamental constituents are used to compute the effects of inter-particle interactions ab initio. The grand Gibbs potential as a function of activities is used to calculate the specific thermodynamic quantities. All other current equations of state considered here (CEFF, FreeEOS, SAHA-S) are based on the free-energy-minimization approach in the so-called ``chemical picture''. The chemical picture aims at providing the detailed ion distributions needed for spectroscopy and opacity calculations, as well as a thermodynamically consistent EOS, for a mixture of many ``almost-ideal'' reacting components. Within the same physical assumptions, both approaches should provide the same thermodynamics description. However, some differences are always observed in practice. 

In the weakly non-ideal approximation, a contribution of Coulomb inter-particle interactions can be added to the thermodynamic potential as a separate term obtained from the results of an external calculation.
In the early EOS formalisms, such as FreeEOS, CEFF and MHD EOS, the Coulomb correction was added in the form of a simplified (linear) Debye-Hückel model or some generalization of it. In more advanced approaches, the Debye-Hückel contribution is obtained as a result of the perturbation procedure in the virial expansion and sequential generation of an activity expansion \citep{Rogers73} as used in physical picture of the OPAL EOS. The same ring-approximation of Debye-Hückel term has been used in the chemical picture of the SAHA-S EOS \citep{Gryaznov04}.

Several additional thermodynamic corrections are needed for precise astrophysical modelling, such as radiative pressure, electron degeneracy, relativistic corrections, ... They are treated as additional terms or extra-corrections in both physical and chemical approaches. The SAHA-S equations of state are the most recent chemical-picture equations of state and they incorporate all principal results from quantum-statistical mechanics including Coulomb corrections and partition functions. This leads to the expectation that differences between the SAHA-S and OPAL equations of state should be minimal in our studies.

\begin{figure}
	\centering
		\includegraphics[width=9cm]{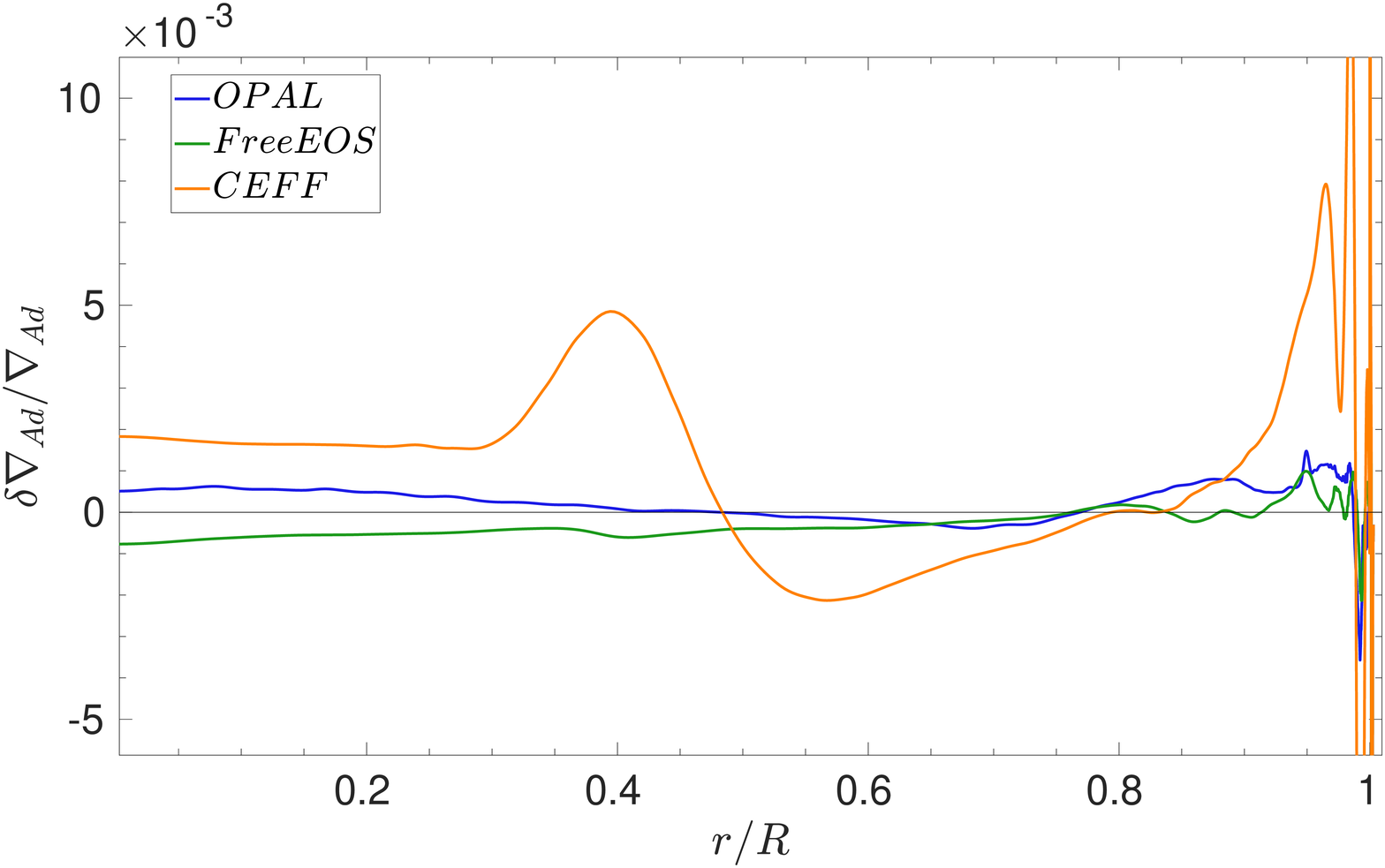}
	\caption{Relative differences in adiabatic gradient between the CEFF, OPAL and FreeEOS equation of state with respect to the SAHA-S equation of state, taken here as reference. $\delta \nabla_{Ad}/\nabla_{Ad}=(\nabla^{EOS}_{Ad}-\nabla^{SAHA-S}_{Ad})/\nabla^{SAHA-S}_{Ad}$, with $EOS$ denoting either CEFF, FreeEOS or the OPAL equation of state.}
		\label{FigNablaAd}
\end{figure}

To illustrate the differences between the various equations of state, we plot in figure \ref{FigNablaAd} an example of differences in thermodynamic quantities, namely the relative differences of the adiabatic gradient obtained with each equation of state from the input profiles of the density, temperature, hydrogen and heavy element abundances, assuming an AGSS09 chemical mixture. Again, the CEFF equation of state is showing divergences when compared to all other equations. The reasons for these differences have not been carefully studied yet, but these discrepancies can be seen already at the beginning of the main sequence and they certainly influence the evolution of the BCZ and thus the settling rates of chemical elements in the models. 

As an additional test, we computed models with similar initial parameters with each equation of state to gain insight on what could be causing the observed differences. This analysis further shows that the OPAL and SAHA-S models were very similar, as both evolutionary sequences lead to similar parameters of the convective zones for both models. We could also observe that the FreeEOS model always showed a higher position of the envelope throughout the evolution and thus leads to a more effective diffusion. This lowers the hydrogen abundance in most of the stellar layers, which implies a lower opacity and thus a higher luminosity at the solar age. To recover the solar luminosity in a calibration, a higher metallicity is required. Since $Z/X$ is imposed for the calibration, $X$ must also be slightly increased and the resulting helium abundance is lower. As for the CEFF model, the behaviour of the settling rate is more complicated as the properties of the convective envelope are very different, probably as a consequence of the discrepancies in thermodynamic quantities which in turn influence the mean molecular weight gradient in the radiative zone and hence the luminosity. This last deduction remains however speculative and further dedicated studies are required to pinpoint the origin of these differences and what they imply for the solar modelling problem. 

\subsubsection{Changing diffusion and convection}
Small variations can also be seen when using the \citet{Paquette} coefficients for diffusion. It appears that they lead to a slightly larger disagreement with the Sun, although not as significant as the one using the CEFF equation of state. The effect is a direct consequence of the fact that the \citet{Paquette} coefficients lead to a less efficient transport of the chemical elements, which induces a slightly lower contrast in chemical composition between the radiative and convective zones which also leads to a less steep temperature gradient. Both these effects imply a slightly higher entropy plateau and small changes in $c^{2}$ and $A$ just below the convective envelope. 

The physical origin of the differences between the \citet{Thoul} and \citet{Paquette} formalism, as presented in this study, is that for the latter case, the diffusion coefficients are computed from the collision integrals using screened Coulomb potentials, whereas the approach of \citet{Thoul} used a cut-off length for the Coulomb interaction placed at the Debye length. In their paper, \citet{Paquette} also argue that, in addition to be more realistic, screened Coulomb potentials are used to mimic to some extent multiparticle collisions.

Similarly, the model computed using the FST formulation of convection, denoted AD1-Free-OPAL-FST in Figs \ref{FigSoundSpeedSTD}, \ref{FigEntropyStd}, and \ref{FigLedouxSTD}, shows very similar results to the MLT model. The only difference is found in the behaviour of the entropy proxy plateau, which is flatter than with the MLT formulation and hence in better agreement with physical expections. This seems to imply that the FST formulation, which considers a whole spectrum of sizes for the convective elements, shows a better agreement with the entropy gradient in the surface region of the Sun and that the resulting surface effect could be smaller. Similar behaviours have been observed with MLT models complemented in the layers close to the surface by a $\mathrm{T}(\tau)$ law from \citet{Vernazza}. Further tests with models patched to averaged hydrodynamical simulations and various boundary conditions are required to check this behaviour to see if other contributors to the surface effect, such as the non-adiabaticity of the pulsation frequencies, could also induce similar trends.

\subsubsection{Changing the metal mixture}\label{secMetalMix}
The uppper-right panel of Figures \ref{FigSoundSpeedSTD}, \ref{FigEntropyStd} and \ref{FigLedouxSTD} illustrate the effects of abundance changes in standard solar models. The GN$93$ and GS$98$ models illustrate the agreement obtained with previous photospheric abundance tables derived from $1-\mathrm{D}$ empirical atmospheric models. From the adiabatic sound speed inversions, the depiction of the problem is quite straightforward, with the large discrepancies resulting from the decrease in heavy elements abundances which strongly reduces the opacity in the radiative zone. From the entropy proxy and Ledoux discriminant inversions, other features start to appear, such as the small deviation of these models around $0.65$ solar radius that is not present in models computed with the recent abundances. As noted in \citet{BuldgenA}, this small deviation is the signature of a slightly too steep temperature gradient at this depth and perhaps a hint at the fact that the metallicity is indeed too high in these models\footnote{However, it should be noted that efficient chemical mixing can lead to a less steep temperature gradient in these regions and that this feature could be erased with an ad-hoc modification of the models.}. In addition to the GN$93$ and GS$98$ abundance tables, we also plot the inversion results for models computed using the meteoritic abundances for all elements except for the volatile elements C, N, O, Ne and Ar, as in \citet{SerenelliComp}, denoted AGSS$09m$. Using this modified abundance table, we find a slightly larger disagreement than with the ``standard'' AGSS$09$ table, similarly to their paper. Finally, the modified abundance table, denoted here AGSS$09Ne$, which includes the $40\%$ increase in Neon over Oxygen determined independently by \citet{Landi} and \cite{Young} seems to perform much better. This is not surprising since Neon is the third contributor to the opacity after oxygen and iron at the BCZ \citep[e.g.][]{Basu08, BlancardOpacDetail}. However, it seems that including this Neon increase generates a similar behaviour to that of the GS$98$ models, which indicates that the steepening of the temperature gradients induced by the Neon increase must be somehow mitigated. The idea of compensating for the decrease in Oxygen through an increase in Neon had already been presented in multiples studies \citep[see e.g.]{Antia05, Basu08, Zaatri2007}, although the increase was pushed to $10\times$ the value found in recent studies and was adjusted such as to recover the agreemeent between AGSS$09$ models and helioseismic constraints. We will come back to this in Section \ref{SecExtraMix}. Moreover, the heavy elements abundance tables have their own uncertainties which lead to an overall uncertainty of around $10\%$ on the solar metallicity value (N. Grevesse, personal communication). This could of course significantly affect the solar modelling problem and emphasizes the need to constantly improve the precision of these measurements, as they are a key ingredient for solar and stellar modelling.

\section{Modified Solar Models}\label{SecNonStdModels}

In addition to standard solar models, we also analysed the changes in the various inverted profiles that could be obtained when applying ad-hoc modifications to some key ingredients of the models. From \citet{Nahar} and \citet{Pradhan}, an increase in the mean Rosseland opacity cannot be excluded from future calculations. Additional experimental results from the Lawrence Livermore National Laboratory are in preparation and could independently confirm the results of the Sandia Z-pinch measurements \citep{Bailey} of the spectral opacity of iron. This increase of the mean Rosseland opacity was recently estimated to be of $9\%$ at a temperature of $\log T \approx 6.3$ in preliminary calculations but higher values could be expected from more accurate computations \citep{Zhao}. The exact origin of these opacity underestimations may stem, at least partially, from inaccurate atomic data for certain key contributors such as iron, silicium, sulfur and to a lesser extent magnesium, oxygen and neon.  In addition to the opacity problem, we found in our previous study that the discrepancies in the Ledoux discriminant observed in the tachocline could be reduced by adding a diffusive extra mixing supposed to mimic the effects of turbulence in a very thin region below the formal Schwarzschild boundary \citep{Gabriel97,Brun99,Brun02}. Besides the chemical mixing, the behaviour of the temperature gradient, as it changes from the adiabatic temperature gradient to the radiative one, is also a source of uncertainty. The modelling of the overshooting region at the BCZ can be treated in various ways, none fully satisfactory, using a diffusive or instantaneous mixing that considers the temperature gradient to be adiabatic or radiative.  In the next sections, we will test the impact of both opacity modifications and the impact of extra-mixing below the formal Schwarzschild boundary on the inversion results of $c^{2}$, $S_{5/3}$ and $A$. We will further comment on the overshooting problem in Sect. \ref{SecExtraMix} and discuss how additional seismic constraints can be used to further constrain the temperature gradient transition and the type of mixing at the BCZ.

\subsection{The opacity problem}\label{SecOpacities}

\begin{figure}
	\centering
		\includegraphics[width=8.5cm]{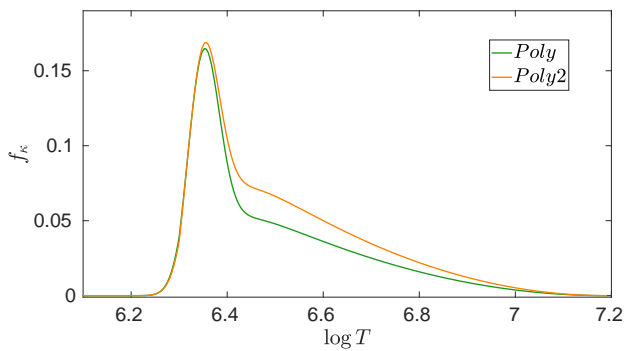}
	\caption{Modification to the opacity profile used in the solar models denoted as ``$Poly$'' (Green) and ``$Poly2$'' (Orange) in Fig \ref{FigOpacPoly} against $\log T$. $f_{\kappa}$ is the increase in relative opacity applied during the evolution (see Eq. \ref{EqOpacForm} and the enclosed discussion).}
		\label{FigOpacModif}
\end{figure}

The main issue with testing changes in the opacity profile is that the degrees of freedom of the solar problem increase tremendously, as one can consider any type of changes which can have various impacts on the models. We limited ourselves to changes motivated by physical considerations and discussions with Prof. Anil Pradhan. First, we started by investigating the impact of a Gaussian increase at various temperatures near the BCZ, using the AGSS$09$ abundance tables, the FreeEOS equation of state and the OPLIB opacity tables as reference ingredients. Each model was recalibrated using a modified mean Rosseland opacity to grasp the effects over the evolution. The modification to the opacity profile is applied as a relative increase factor to the mean Rosseland opacity as follows
\begin{align}
\kappa^{'}=(1+f_{\kappa})\kappa, \label{EqOpacForm}
\end{align}
with $\kappa$ the original mean Rosseland opacity, $f_{\kappa}$ the function defining the opacity modification and $\kappa^{'}$ the resulting modified opacity which is applied similarly during the whole evolution. At first, we start with a very narrow Gaussian increase for $f_{\kappa}$ with 
\begin{align}
f_{\kappa}=A\exp{-\frac{(\log T - \log T_{\mathrm{Ref}})^{2}}{2\Delta^{2}}},
\end{align}
with $A$ the amplitude of the increase, $\log T_{\mathrm{Ref}}$ the peak temperature of the Gaussian and $\Delta$ its standard deviation. More complex modifications will also be considered (see Fig. \ref{FigOpacModif}) and applied similarly to the original mean Rosseland opacity profile.

\begin{figure*}
	\centering
		\includegraphics[width=17cm]{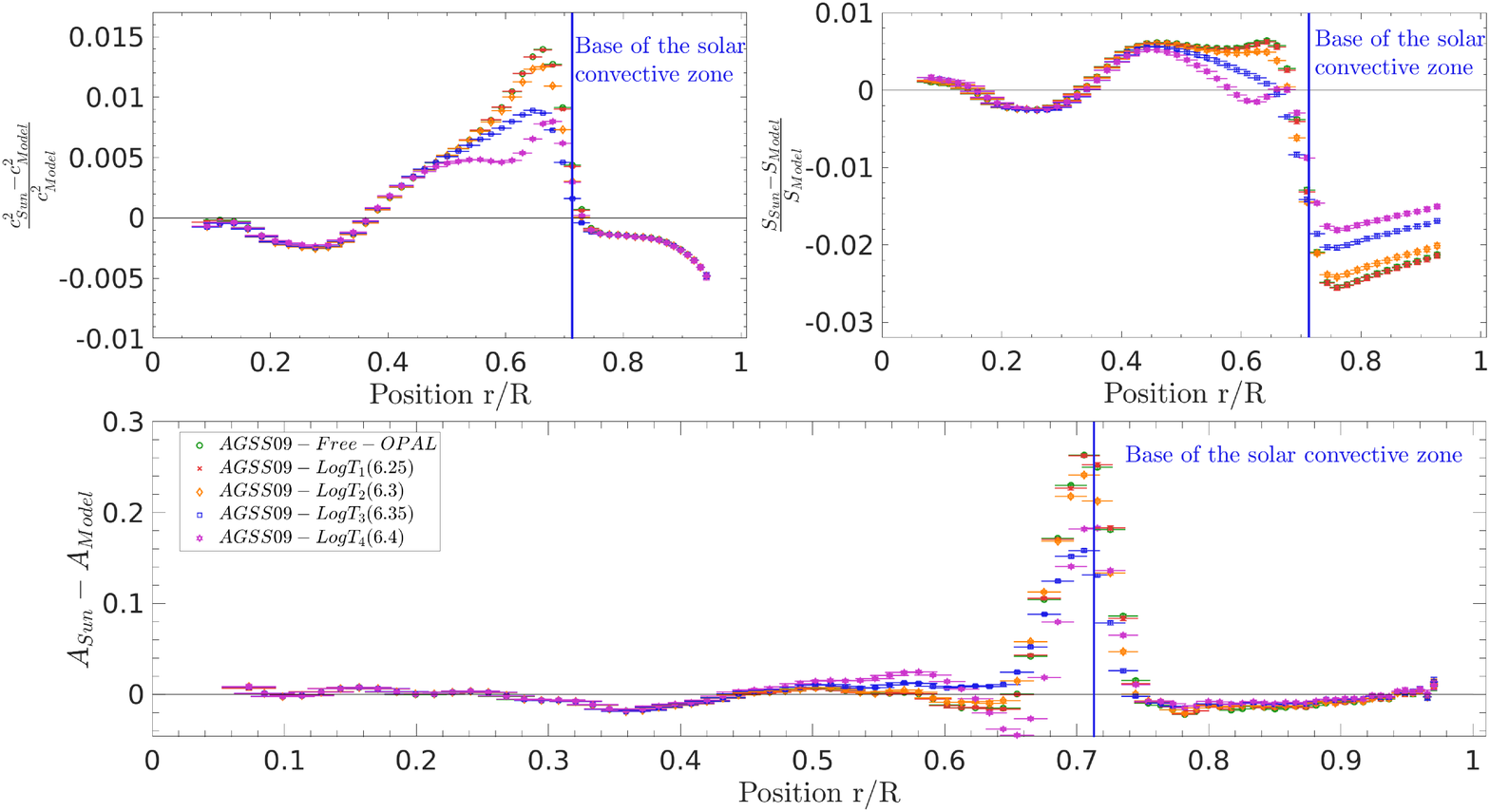}
	\caption{Inversion results for models with modified opacities including a gaussian peak of fixed height and width at various temperatures (see text for details). Upper-left panel: relative differences in squared adiabatic sound speed. Upper-right panel: relative differences in entropy proxy, $S_{5/3}$. Lower panel: differences in Ledoux discriminant, $A$.}
		\label{FigOpacT}
\end{figure*}

We tested four temperatures for the peak of the Gaussian increase, namely $\log T_{\mathrm{ref}} =6.25$, $6.30$, $6.35$ and $6.4$ (denoted hereafter as $T_{1}$, $T_{2}$, $T_{3}$ and $T_{4}$, respectively) with maximum amplitude coefficients of $10\%$, decreasing down to $3\%$ at $0.05$ dex from the center of the Gaussian. The results are shown in Fig. \ref{FigOpacT} and prove that the location of the peak has a large impact on the inverted profiles. At $\log T_{1} =6.25$ (red), the peak has almost no effect. Closer to the BCZ, at $\log T_{2} =6.3$ (orange), the models start to show a slightly better agreement for all quantities. At $\log T_{3} =6.35$ (blue), the changes are drastic, with a large decrease in the height of the entropy plateau, changes in adiabatic sound speed and in the convective parameter at the BCZ. When the peak is moved to $\log T_{4} =6.4$ (purple), a slight improvement is observed for the entropy plateau and for the sound speed profile, but additional deviations are seen in the profile of the Ledoux discriminant around $0.65$ solar radii. From \citet{BuldgenA}, we could identify that this results from a too steep temperature gradient, which will not be seen in the $S_{5/3}$ and $c^{2}$ inversions. It is also worth noticing the similarities between the profiles of this model and that of a GN$93$ model built with the OPAL opacities, illustrating the degeneracy of the solar modelling problem in terms of abundances and opacities. We conclude that the optimal positioning for the opacity peak is at $\log T=6.35$, corresponding to the temperature of an iron opacity peak, which could be the source of the discrepancies.

\begin{figure*}
	\centering
		\includegraphics[width=17cm]{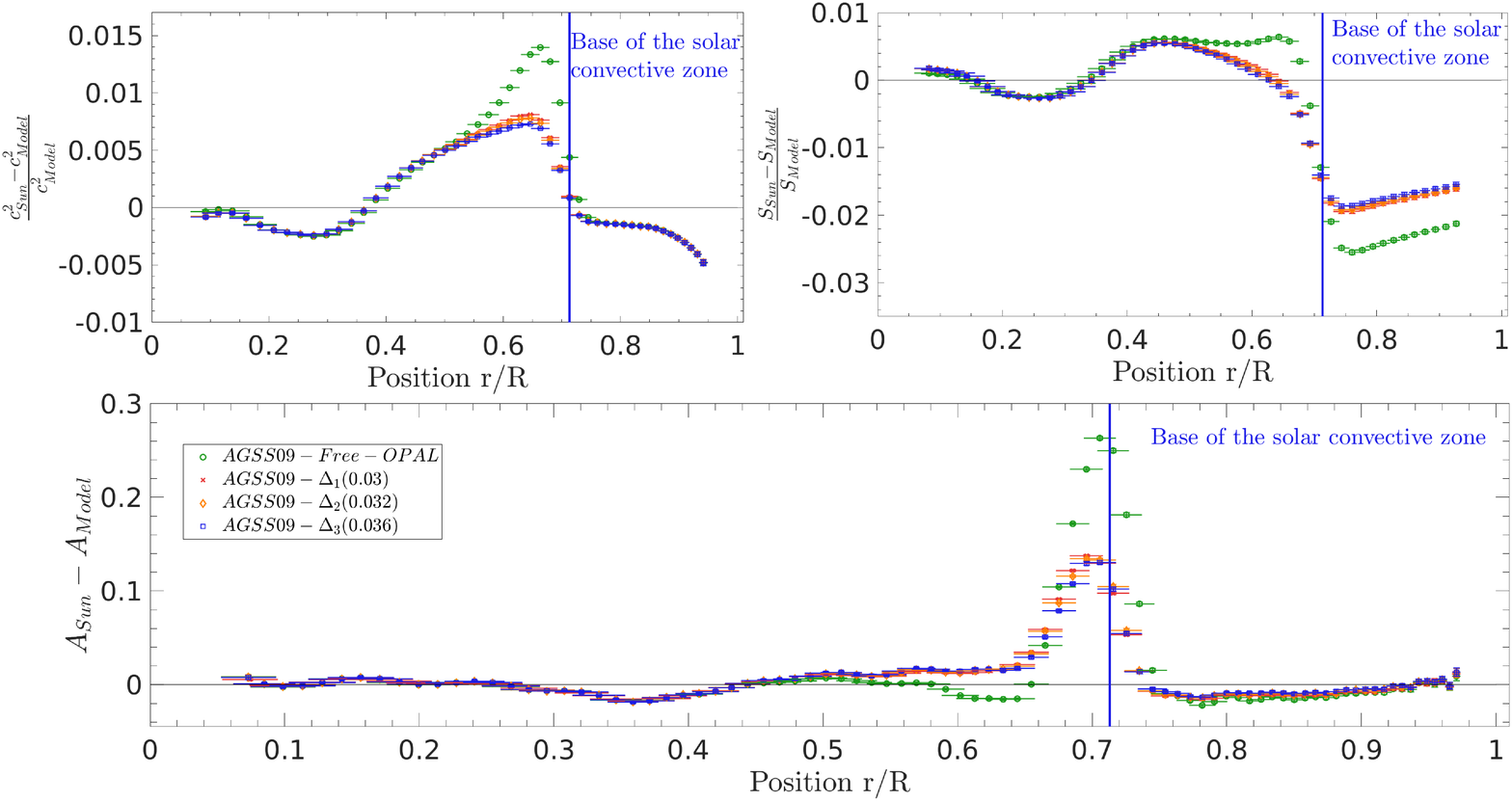}
	\caption{Inversion results for models with modified opacities including a gaussian peak at a fixed temperature with different widths but the same height (see text for details). Upper-left panel: relative differences in squared adiabatic sound speed. Upper-right panel: relative differences in entropy proxy, $S_{5/3}$. Lower panel: differences in Ledoux discriminant, $A$. }
		\label{FigOpacDelta}
\end{figure*}

Besides the position in temperature, we tested the importance of the width of the Gaussian increase in opacity. We considered a position for the peak at $\log T=6.35$ and a maximal increase of $13\%$ for all models shown in Fig. \ref{FigOpacDelta}. We started with $\Delta_{1}\approx 0.03$ (red), which allows for a decrease down to $3\%$ at $0.05$ dex from the peak, then slightly modified the decrease with $\Delta_{2} \approx 0.032$ (orange) and $\Delta_{3} \approx 0.036$ (blue). These values imply that at $0.05$ dex from the peak temperature, the opacity change the opacity increased by $3.6\%$ and $4.4\%$, respectively. These variations are intentionally quite small and as expected, do not have a large impact on the models, with the exception of $\Delta_{3}$ which starts showing slight differences that can arguably not be inputed to numerical uncertainties. It is however interesting to notice that none of the models have led to an improvement in the convective zone helium abundance with respect to the threshold value of $0.245$ we chose, while some can significantly reduce the discrepancies in the position of the BCZ, the height of the entropy plateau and the sound speed profile compared to the standard AGSS$09$ model (see the first line of table \ref{tabSTDModels}).

\begin{figure*}
	\centering
		\includegraphics[width=17cm]{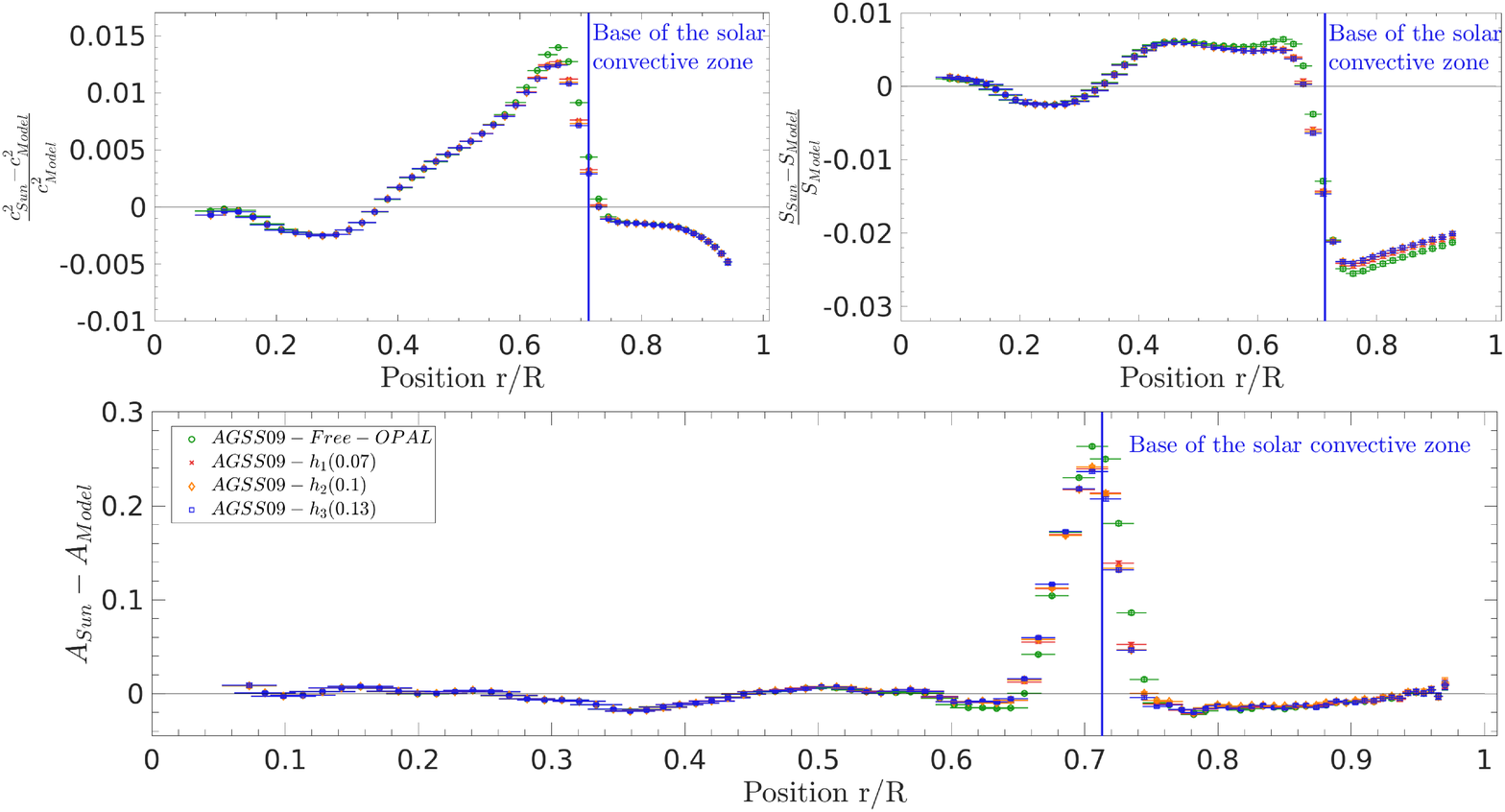}
	\caption{Inversion results for models with modified opacities including a gaussian peak at a fixed temperature with different heights and widths such as to keep the same increase at a fixed distance of the maximum (see text for details). Upper-left panel: relative differences in squared adiabatic sound speed. Upper-right panel: relative differences in entropy proxy, $S_{5/3}$. Lower panel: differences in Ledoux discriminant, $A$.}
		\label{FigOpach}
\end{figure*}

Similarly, we tested the impact of the maximum height of the Gaussian increase in opacity. We considered a position at $\log T=6.3$ and analysed the importance of increasing the height of the peak while simultaneously reducing its width to ensure that its amplitude was reduced to $3\%$ at $0.05$ dex from the peak temperature. We considered a height of $7\%$, $10\%$ and $13\%$ for the peak. The models are denoted respectively $h_{1}$ (red), $h_{2}$ (orange) and $h_{3}$ (blue) in table \ref{tabOPACModels}. From figure \ref{FigOpach} and table \ref{tabOPACModels}, we see that the changes are minimal. Significant changes are only seen for the position of the BCZ, which is affected by the local steepening of the temperature gradient. This limited effect was already seen in Fig. \ref{FigOpacT} where one could see that at $\log T=6.25$ and $\log T=6.3$ the very localised opacity increase we considered had almost no effect. This is further confirmed in our tests when varying the height of the peak. However, this limited impact is only valid for such very localized changes and, as we will see below, when more extended modifications are considered, the region at $\log T= 6.3$ plays a crucial role. 

Besides localized increases, we tested a slightly more extended variation of the opacity, shown in Fig. \ref{FigOpacModif}. The explanation behind this modification is that most of the uncertainties on the opacities reside in the iron peak and around it, at $\log T=6.35$. However, due to the rapid increase in higher ionization states for most of the elements, the opacity computation are expected to be more robust as photon absorption becomes much less significant. Thus, we considered that the opacity difference would rapidly decrease with increasing temperature. We denoted models computed with this modification to the mean Rosseland opacity as ``$Poly$'' in the following figures, tables and discussions. The behaviour of this profile is however not by any means accurate and should only be seen as a qualitative estimate of what an opacity modification resulting from more accurate calculations could be. We did not apply any modification in the opacity profile at lower temperatures since they would be in the convective zone of the model, hence the sharp decrease in Fig. \ref{FigOpacModif}. However, this does not mean that such modifications are not expected\footnote{They could potentially be larger than $10\%$.} and could not have an impact for other stars than the Sun.

\begin{figure*}
	\centering
		\includegraphics[width=17cm]{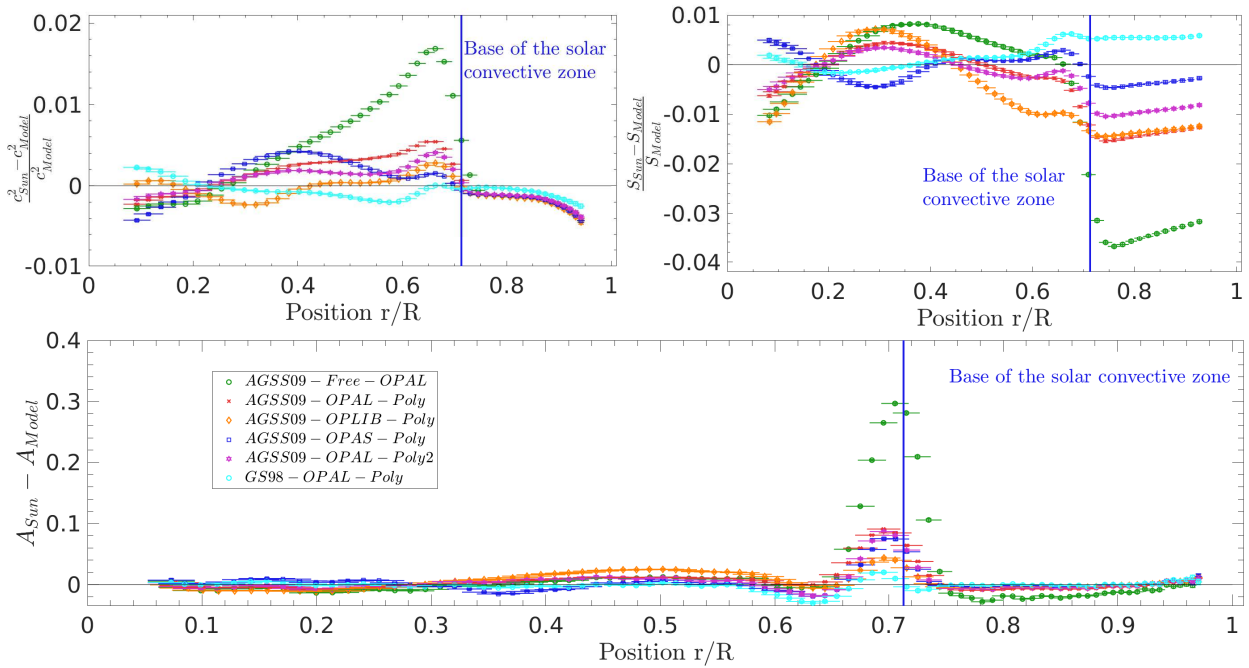}
	\caption{Inversion results for models with modified opacities including a two-components modification for the opacity (see text for details). Upper-left panel: relative differences in squared adiabatic sound speed. Upper-right panel: relative differences in entropy proxy, $S_{5/3}$. Lower panel: differences in Ledoux discriminant, $A$.}
		\label{FigOpacPoly}
\end{figure*}

The changes brought by this opacity modification are quite impressive, showing for all models a drastic improvement of the height of the entropy plateau, of the sound speed and Ledoux discriminant profiles, as can be seen in Fig. \ref{FigOpacPoly}. It is interesting to see that a small change around $3\%$ between $0.4$ and $0.5$ solar radii has such a drastic effect on the sound speed profile. Moreover, this opacity modification has a non-negligible impact on the helium abundance in the convective zone, because a more widespread increase leads to a lower initial hydrogen abundance required to reproduce the solar luminosity at the solar age. We also note that our opacity modification has a similar amplitude than that of \citet{JCD2010,Ayukov2011} (see the dashed curve of Fig. $13$ in the former and Fig. $4$ in the latter) at the BCZ, but with a much steeper decrease at higher temperature \citep[See also][for a study in terms of opacity kernels.]{JCD98}. Moreover, modifications of about $3\%$ are consistent with differences between various opacity tables \citep[e.g.][]{Guzik05, Guzik06}, which can be used to estimate their ``optimistic'' uncertainties. At the level of accuracy of helioseismology, various hypotheses in the formalism of microscopic diffusion and variations in the equation of state can also induce differences between various modellers \citep[see e.g.][ and references therein for further discussions.]{Montalban06}. Typically, uncertainties on the diffusion velocities for iron and oxygen can reach values of about $35\%$ resulting from assuming full ionization instead of partial ionization of the stellar material. Effects of radiative accelerations should remain small, but not fully negligible for iron \citep{Turcotte,Gorshkov10}, especially if the opacity of this element is underestimated. 
\begin{table*}[t]
\caption{Parameters of the solar models with modified opacity used in this study}
\label{tabOPACModels}
  \centering
  \resizebox{\linewidth}{!}{%
\begin{tabular}{r | c | c | c | c | c | c | c | c | c | c}
\hline \hline
\textbf{$\left(r/R\right)_{BCZ}$}&\textbf{$\left( m/M \right)_{CZ}$} &\textbf{$Y_{CZ}$}&\textbf{$Z_{CZ}$}&\textbf{$Y_{0}$}&\textbf{$Z_{0}$}&\textbf{EOS}&\textbf{Opacity}&\textbf{Abundances} & \textbf{Diffusion} & \textbf{Convection}\\ \hline
$0.7200$&$0.9776$&$0.2300$&$0.01372$& $0.2587$&$0.01519$ & FreeEOS & OPLIB+$\log T_{1}$ $(6.25)$ & AGSS09 & Thoul & MLT\\
$0.7165$&$0.9769$&$0.2302$&$0.01371$&$0.2586$ &$0.01518$ & FreeEOS & OPLIB+$\log T_{2}$ $(6.30)$ & AGSS09 & Thoul & MLT\\
$0.7155$&$0.9766$&$0.2304$&$0.01371$& $0.2586$&$0.01515$ & FreeEOS & OPLIB+$\log T_{3}$ $(6.35)$ & AGSS09 & Thoul & MLT\\
$0.7195$&$0.9773$&$0.2303$&$0.01371$&$0.2588$ &$0.01517$ & FreeEOS & OPLIB+$\log T_{4}$ $(6.40)$ & AGSS09 & Thoul & MLT\\
$0.7139$&$0.9768$&$0.2305$&$0.01368$& $0.2585$&$0.01514$ & FreeEOS & OPLIB+$\Delta_{1}$ & AGSS09 & Thoul & MLT\\
$0.7136$&$0.9762$&$0.2301$&$0.01372$& $0.2585$&$0.01514$ & FreeEOS & OPLIB+$\Delta_{2}$ & AGSS09 & Thoul & MLT\\ 
$0.7134$&$0.9761$&$0.2301$&$0.01363$& $0.2585$&$0.01514$ & FreeEOS & OPLIB+$\Delta_{3}$ & AGSS09 & Thoul & MLT\\ 
$0.7174$&$0.9771$&$0.2301$&$0.01370$&$0.2585$&$0.01518$ & FreeEOS & OPLIB+$h_{1}$ & AGSS09 & Thoul & MLT\\
$0.7165$&$0.9769$&$0.2302$&$0.01371$&$0.2586$ &$0.01518$ & FreeEOS & OPLIB+$h_{2}$ & AGSS09 & Thoul & MLT\\
$0.7158$&$0.9761$&$0.2302$&$0.01371$&  $0.2587$&$0.01518$ & FreeEOS & OPLIB+$h_{3}$ & AGSS09 & Thoul & MLT\\
$0.7124$&$0.9758$&$0.2394$&$0.01355$& $0.2683$&$0.01496$ & FreeEOS & OPAL+Poly & AGSS09 & Thoul & MLT\\
$0.7104$&$0.9750$&$0.2332$&$0.01366$& $0.2608$&$0.01506$ & FreeEOS & OPLIB+Poly & AGSS09 & Thoul & MLT\\
$0.7089$&$ 0.9751$&$0.2354$&$0.01362$&$0.2634$& $0.01503$ & FreeEOS & OPAS+Poly & AGSS09 & Thoul & MLT\\
$0.7118$&$0.9755$&$0.2404$&$0.01353$& $0.2691$& $0.01493$ &FreeEOS & OPAL+Poly2 & AGSS09 & Thoul & MLT\\
$0.7056$&$0.9736$&$0.2495$&$0.01694$& $0.2791$&  $0.01885$&FreeEOS & OPAL+Poly & GS98 & Thoul & MLT\\
\hline
\end{tabular}
}
\end{table*}

Observing such changes from the introduction of a rather small additional component in the ad-hoc increase of the opacity below the iron opacity peak around $\log T =6.35$ is surprising. In truth, it is due to changes in the opacity derivatives. The introduction of a localized increase unsurprisingly only impacts the models locally, while the addition of the polynomial decrease allows the Gaussian peak to impact the opacity profile at wider depth. Similarly, a constant increase of the opacity at all temperatures does not strongly improve the properties of the models, except the helium abundance, because the temperature gradient is sensitive to the whole landscape of the opacities and thus to their derivatives \citep[we note that this is in agreement with][]{Ayukov2011, Colgan, Guzik16,Ayukov2017}. These variations of the solar calibrated models with respect to various opacity modifications emphasize well the advantage of taking into account multiple constraints and also illustrates the very constraining nature of the determination of the helium abundance in the solar convective zone.

All models built with the AGSS09 abundances show an agreement in sound speed with the Sun very similar to that of the ModelS from \citet{MODELS}. Moreover, looking at fig $6$ from \citet{Kosovichev}, the models built with the OPAL (red) and OPAS (blue) opacities show a similar agreement in the terms of Ledoux discriminant than ModelS. However, none of them has a satisfactory helium abundance and height of the entropy plateau. This illustrates the advantage of combining multiple inversions and show that the inversion of the sound speed alone can hide some compensation effects.

Such effects are illustrated by the model denoted $Poly2$ (purple), which considers a slightly slower decrease of the opacity modification with temperature, and the model built with the OPLIB opacities (orange) and the $Poly$ modification in Fig. \ref{FigOpacPoly}. As can be seen, both these models start inducing a slight disagreement around $0.65$ solar radii in A. This effect is also present in ModelS, due to the higher metallicity but here it is a result of the opacity modification introduced in both models. 

For the sake of comparison, we tested the extended opacity modification in a model using the GS98 abundances (light blue). We see that this model is in disagreement with helioseismic constraints. Indeed, it shows a very deep BCZ, quite large discrepancies in Ledoux discriminant and sound speed around $0.6$ solar radii and a significant discrepancy in entropy proxy just below the convective zone, as can be seen from Fig. \ref{FigOpacPoly}. Moreover, it shows a quite low entropy proxy plateau, which is a clear sign of a too steep temperature gradient below the BCZ. However, the helium abundance in this model is in very good agreement with helioseismology and the discrepancies in the deeper radiative region, between $0.2$ and $0.5$ solar radii, in the inverted profiles are very small. This confirms that if an increased opacity is obtained in updated opacity tables, the higher metallicity solar abundances will disagree with helioseismology.

As for the models built with the AGSS09 abundances, we note that the OPLIB opacities induce a steeper temperature gradient near the BCZ despite their overall lower values in the solar radiative zone. Consequently, the model including the $Poly$ modification built with the OPLIB opacities has a steeper temperature gradient near the envelope than the $Poly$ model built with the OPAL opacities. Regarding the model $Poly2$, the steeper temperature gradient is a consequence of the higher amplitude of the changes made in the opacity profile near the BCZ. As a matter of fact, both models closely resemble ModelS, with the exception of their lower helium abundance in the envelope. This very low abundance is particularly problematic for the OPLIB and OPAS opacities which, despite an increase in opacity over a broad domain, do not seem to be able to reproduce this constraint. This could indicate that they present a significantly larger underestimation of the opacity in the radiative region than other tables. Similarly, one can note that the model built with the OPAS opacity tables also shows larger discrepancies in the deep radiative region than the models built with the OPAL tables, and a too deep BCZ. From a theoretical point of view, one expects a certain amount of overshoot which would extend the region in which the temperature gradient is adiabatic. This, in turn, implies here an even deeper limit of the convective zone, which would disagree even more with the helioseismology. 

Combining all these analyses together confirms that no opacity tables seem to provide a satisfactory agreement with the Sun if the AGSS$09$ mixture is used. However, this also seems to indicate that changing the opacity only near the BCZ will be insufficient to solve the solar modelling problem. Extending the opacity increase towards higher temperature can improve the agreement, but other constraints, such as the position of the BCZ and the helium abundance in the convective zone have to be taken into account in the overall analysis. These constraints indicate that none of the models presented in this section are in satisfactory agreement with the Sun. This is especially problematic for the models computed with the OPAS and OPLIB tables, which are the latest generations of opacity tables. One way to partly solve this problem is to consider a constant increase in opacity of about $5\%$ over the whole solar structure, which is still close to the differences found between various former standard tables \citep{Guzik05, Guzik06}. Such a modification does not largely impact the profile inversions nor the position of the BCZ, but can strongly increase the helium abundance by about $0.006$, which would bring the OPAS models back in a range of helium values more acceptable (although still quite low).

\subsection{Impact of additional mixing and abundance changes}\label{SecExtraMix}

Besides uncertainties on opacities, standard solar models are also lacking a proper representation of the tachocline region, where additional mixing of the chemical elements is supposed to occur \citep{Brun99, Brun02} and where the transition from the adiabatic to the radiative temperature gradient occurs in a smoother way than in models \citep{SpiegelZahn1992,Monteiro94,Xiong01,Rempel04,Hughes2007,YangI, YangII, JCDOV,Zhang14, Hotta}.  

In a previous study \citep{BuldgenA}, we showed that adding a localized additional diffusive mixing below the BCZ could reduce the discrepancies in the Ledoux discriminant inversions. Consequently, we decided to test models where one would add extra-mixing in addition to the extended opacity modification presented in Sect. \ref{SecOpacities}. The behaviour and intensity of this mixing should be further investigated. Our results only depict the qualitative behaviour to be expected if such an extra-mixing is included. Comparisons with hydrodynamical simulations \citep[Such as in][]{Viallet} and models reproducing the solar rotation profile \citep{Charbonnel,Eggenberger} could provide guidelines to empirical approaches for producing a smoother profile of the mean molecular weight, expected at the BCZ. On a longer timescale, improving the modelling of convection in stellar interiors altogether is the true concern of such studies. Currently, besides the opacity tables, the efficiency of chemical mixing in this region is the largest contributor to the uncertainties \citep{Vinyoles}.

For the tests presented here, we added the extra-mixing below the envelope as presented in \citet{BuldgenA}, using a turbulent diffusion coefficient of the form
\begin{align}
D_{Turb}=D\left( \frac{\rho_{\mathrm{cz}}}{\rho(r)} \right)^{N}, \label{eqDTURB}
\end{align}
with the free parameters $D$ and $N$ fixed respectively to $50$ and $2$ for the model AGSS$09$Ne-Poly-DTurb and $7500$ and $3$ for the model AGSS$09$Ne-Poly-DTurb-Prof. In Eq. \ref{eqDTURB}, $\rho_{\mathrm{cz}}$ is the density value at the BCZ. The values of AGSS$09$Ne-Poly-DTurb-Prof are based on the models of \citet{Proffitt} and \citet{Richard2005} that reproduced the lithium depletion in solar models \citep[see also][for similar investigations using sound speed and lithium and beryllium depletion]{Richard96Sun,Piau2001,Thevenin17}.  

Besides diffusive mixing, we considered models using instantaneous mixing below the BCZ. Model AGSS$09$Ne-Poly-Ov-Rad and AGSS$09$Ne-Poly-Ov-Ad included an extension of the mixed region over $0.3 \mathrm{H}_{p}$, around the order of magnitude in \citet{JCDOV}. In model AGSS$09$Ne-Poly-Ov-Ad, we used an adiabatic temperature gradient in all the overshooting region, while for model AGSS$09$Ne-Poly-Ov-Rad, the temperature gradient was kept to its radiative value. The impact of these changes is discussed below. Other forms of mixing exist \citep[see e.g.]{Zahn92}. However, due to the uncertainties linked to the rotational transport in the Sun, such formalisms would need to be complemented by other mechanisms and further tested, which is beyond the scope of this study.

The extent of the mixed region is of $0.03 \sim 0.04$ solar radii, similar to the estimated width of the tachocline \citep{corbard99,Elliott99Tacho,Hughes2007}. We did not consider an additional overshoot in the models including diffusive mixing, as the opacity increase was sufficient to place the Schwarzschild limit quite low. This confirms that our opacity increase might be too large near the BCZ. We also included the $40\%$ increase in Neon abundance from \citet{Landi} and \citet{Young}, since it also provided a slight improvement. 

\begin{figure*}
	\centering
		\includegraphics[width=17cm]{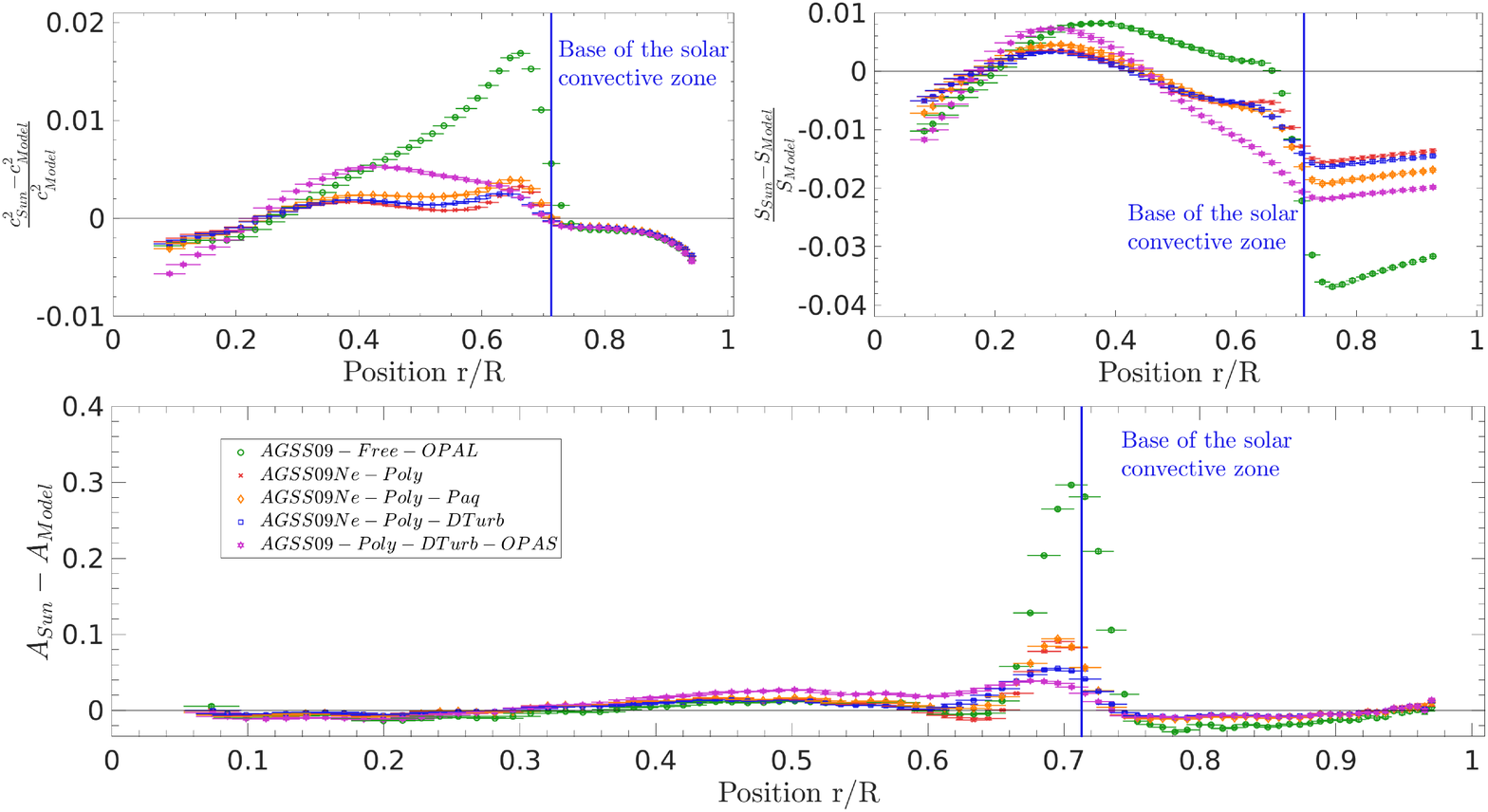}
	\caption{Inversion results for models with modified opacities including a two-components modification of opacity and various formalisms for the mixing of chemical elements (see text for details). Upper-left panel: relative differences in squared adiabatic sound speed. Upper-right panel: relative differences in entropy proxy, $S_{5/3}$. Lower panel: differences in Ledoux discriminant, $A$.}
		\label{FigOpacPolyD4}
\end{figure*} 

\begin{figure*}
	\centering
		\includegraphics[width=17cm]{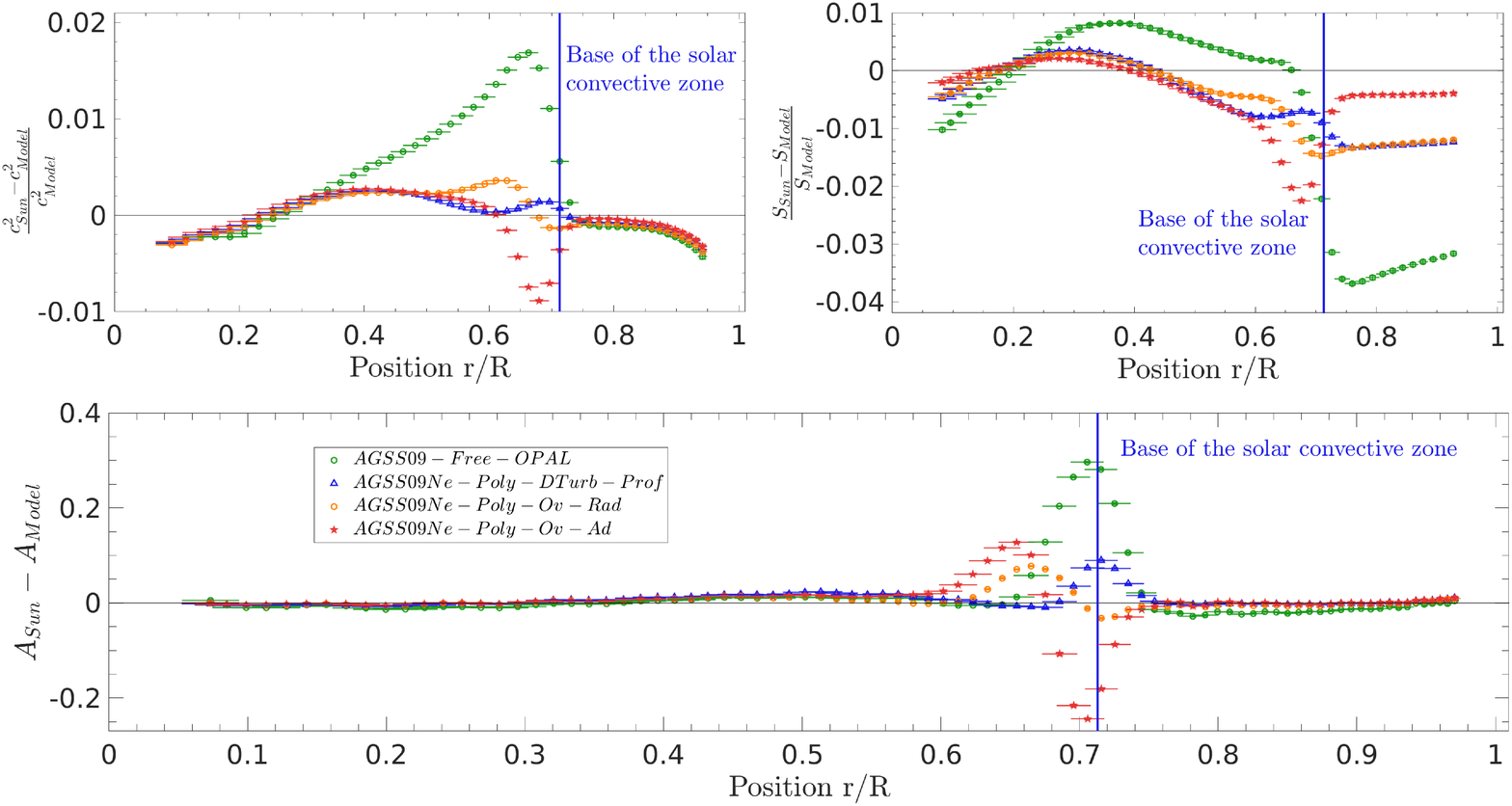}
	\caption{Inversion results for models with modified opacities including a two-components modification of opacity and including overshooting or turbulent diffusion following \citet{Proffitt} for the mixing of chemical elements (see text for details). Upper-left panel: relative differences in squared adiabatic sound speed. Upper-right panel: relative differences in entropy proxy, $S_{5/3}$. Lower panel: differences in Ledoux discriminant, $A$.}
		\label{FigOpacPolyOV}
\end{figure*}

The results of these inversions are presented in Fig. \ref{FigOpacPolyD4} for models using the \citet{Paquette} (in orange), \citet{Thoul} (in red) diffusion coefficients and one model including turbulent diffusion as in \citet{BuldgenA} (in blue) whereas models including radiative or adiabatic overshooting with a fully mixed region (in orange and red, respectively) and the model including the coefficients of \citet{Proffitt} for turbulent diffusion (in blue) are shown in Fig. \ref{FigOpacPolyOV}. Additional informations on each model can be found in table \ref{tabMixModels}. We see from Figs. \ref{FigOpacPolyD4} and \ref{FigOpacPolyOV} that the agreement is further improved with the increase in Neon abundance, both in the inversions and in the helium abundance in the envelope. It should be noted that the increase in helium is also partially due to the use of the SAHA-S equation of state and to the inclusion of turbulent diffusion. Using the \citet{Paquette} diffusion coefficients leads to a slightly higher surface helium abundance but at the expense of a slightly worse agreement of the inversions. 

The models with turbulent diffusion show the best agreement with the Sun, especially, and unsurprisingly, the model using values of \citet{Proffitt} for the parameters of Eq. \ref{eqDTURB}. It has good value for the helium abundance and position of the BCZ while simultaneously showing quite good agreement with inversions. In comparison, the models including overshooting and instantaneous mixing perform poorly. However, the shift of the entropy plateau in the model including adiabatic overshooting illustrates the dependence of this quantity to the transition in temperature gradient just below the BCZ. Yet, while the entropy plateau is quite well reproduced, a glitch is generated deeper in the model, which is in clear contradiction with solar structure. This feature is not present in the model including radiative overshooting, demonstrating that this results from the temperature gradient. However, radiative overshooting appears to be unable to induce a shift in the entropy plateau, while it can reduce the errors in sound speed and Ledoux discriminant. The obvious and well-known conclusion of these observations is that the transition in temperature gradient is improperly reproduced by both models and should be somewhere in between these extremes. It is also worth mentioning that models in Fig. \ref{FigOpacPolyOV} (except the standard model) reproduce the lithium destruction observed in the Sun, while models in Fig. \ref{FigOpacPolyD4} do not.

Ultimately, our tests show the interplay between chemical mixing and opacities but also that the equation of state plays a non-negligible role at the accuracy level we aim for with solar models. Moreover, none of the models are able to fully reproduce each of the profiles and discrepancies are still present in the tachocline. This implies that the transition of the temperature gradient just below the envelope has a significant impact on these disagreements. Hence, our inversions should be coupled with phase shift analysis techniques as in \citet{RoxVor94PhaseShift,MonteiroOne,JCDOV} for an optimal diagnostic.

\begin{table*}[t]
\caption{Parameters of the solar models with modified opacities and additional mixing used in this study}
\label{tabMixModels}
  \centering
  \resizebox{\linewidth}{!}{%
\begin{tabular}{r | c | c | c | c | c | c | c | c | c | c}
\hline \hline
\textbf{$\left(r/R\right)_{BCZ}$}&\textbf{$\left( m/M \right)_{CZ}$} &\textbf{$Y_{CZ}$}&\textbf{$Z_{CZ}$}&\textbf{$Y_{0}$}&\textbf{$Z_{0}$}&\textbf{EOS}&\textbf{Opacity}&\textbf{Abundances} & \textbf{Diffusion} & \textbf{Convection}\\ \hline
$0.7122$&$0.9757$&$0.2416$&$0.01385$&$0.2692$& $0.01494$ & SAHA-S & OPAL+Poly & AGSS09Ne & Thoul & MLT\\
$0.7129$&$0.9761$&$0.2427$&$0.01383$&$0.2678$& $0.01483$ & SAHA-S & OPAL+Poly & AGSS09Ne & Paquette & MLT\\
$0.7106$&$0.9762$&$0.2425$&$0.01383$& $0.2685$ &$0.01466$ & SAHA-S & OPAL+Poly & AGSS09Ne & Thoul+$D_{Turb}$ & MLT\\
$0.7106$&$0.9762$&$0.2374$&$0.01359$& $0.2645$&$0.01490$ & SAHA-S & OPAS+Poly & AGSS09 & Thoul+$D_{Turb}$  & MLT\\
$0.7121$&$0.9756$&$0.2460$&$0.01376$& $0.2696$ &$0.01500$ & SAHA-S & OPAL+Poly & AGSS09Ne & Thoul+$D_{Turb}-\mathrm{Prof}$ & MLT\\
$0.7118$&$0.9757$&$0.2437$&$0.01381$& $0.2692$&$0.01495$ & SAHA-S & OPAL+Poly & AGSS09Ne &  Thoul+$\mathrm{Ov}-\mathrm{Rad}$ & MLT\\
$0.71056$&$0.9751$&$0.2438$&$0.01381$& $0.2700$&$0.01506$ & SAHA-S & OPAL+Poly & AGSS09Ne & Thoul+$\mathrm{Ov}-\mathrm{Ad}$ & MLT\\
\hline
\end{tabular}
}
\end{table*}

A less positive observation is made for the OPAS opacities (in purple in Fig. \ref{FigOpacPolyD4}), which show a low helium abundance and too deep a BCZ, despite the inclusion of extra-mixing. The inversion results still show quite large discrepancies in the radiative region. Comparing figure \ref{FigEntropyStd} and the entropy inversions in Figs. \ref{FigOpacPoly} and \ref{FigOpacPolyD4}, we see that the opacity increase has somewhat improved the agreement in the radiative regions for the OPAL models but has almost no effect on the OPAS models. Similar conclusions are drawn from sound speed inversions in Figs. \ref{FigSoundSpeedSTD}, \ref{FigOpacPoly} and \ref{FigOpacPolyD4}.

Furthermore, the OPAS model is built without the Neon increase, because in their actual form, the OPAS opacities are limited to only one chemical composition. It is thus difficult to properly assess the impact of the changes in abundances and also test how models built with the GS$98$ or GN$93$ composition would behave. Would they show similar problems in the Ledoux discriminant inversions as models built with the OPLIB opacities? 

\section{Conclusion}
	
In this study, we have discussed the potential of combining multiple structural inversions to gain a deeper knowledge of the current inaccuracies of solar models and possible solutions to reconcile helioseismology with the AGSS$09$ abundances. 

First, we presented in Sect. \ref{SecCombinedInversions} the current state of the solar modelling problem using inversions of the sound speed, entropy proxy and Ledoux discriminant for various combinations of standard physical ingredients of solar models. For each combination, we give the position of the BCZ, the mass coordinate at this position and the initial and present-day photospheric abundances of both helium and the heavy elements. We tested various opacity tables, formalisms for diffusion and convection, abundance tables and equations of state. These calibrations show that no combination of current standard ingredient provide a satisfactory agreement for the AGSS$09$ models, but that higher metallicity models are not perfect either. We also find that using the latest opacity tables does not induce a unequivocal improvement. In fact, while the BCZ and the inversion results are improved, this is made at the expense of the helium abundance. This is particularly problematic for the OPLIB opacities.

In Sect. \ref{SecNonStdModels}, we investigated the potential of combining inversions to constrain changes in the opacity profile and chemical mixing at the BCZ to solve the solar modelling problem. We considered modifications to the opacity profile based on physical arguments. We started with localized modifications in the form of a Gaussian peak. We tested the positioning, width and height of these increases and found the optimal positioning of the peak to be at $\log T = 6.35$, in a temperature regime very close to that of an iron opacity peak. At this position, slight increases in the width of the peak had a limited effect on the results but a too broad peak would clearly impact them. Similarly, if the opacity increase is localized and located at $\log T =6.3$, the improvement in the models is quite scarce and variations in the height of the peak did not have a significant impact on the models. These results lend credence to the argument that part of the solar problem is linked to the iron opacity at this regime of physical conditions. 

To further test opacity modifications, we implemented more elaborated variations, including potential smaller opacity underestimations at higher temperatures. While these tests remain purely speculative, they indicate that a relatively moderate opacity modification (of around $15\%$ near the BCZ and $3\%$ at higher temperature) can significantly increase the agreement between low metallicity models and helioseismic results. Furthermore, analyzing simultaneously the inversion results prove that a satisfactory agreement in one quantity can be disqualified by another and thus help disentangle the multiple contributors to the observed discrepancies. This indicates that an approach using simultaneously all the information of these inversions could perhaps be used to derive a more precise estimation of the required opacity profile in the radiative zone, if an equation of state and a chemical composition profile for the model are assumed. 

These tests on the opacity showed that the helium abundance in the convective zone remained quite low for low metallicity models, even if an extended modification is used. Moreover, the BCZ quickly drops below the helioseismic value, implying an inadequate modification of the opacity. Reconciling these constraints with seismic estimates is only possible if a global opacity underestimation is considered instead of the profile we used or if other sources are invoked. Hence, we used models with a modified Neon abundance according to recent estimates, a form of extra-mixing and used the equation of state that could push the results towards a better agreement in terms of helium and BCZ (the SAHA-S equation of state). For these cases, the improvement of the models is far more drastic, the helium abundance is higher due to the higher metallicity, the effects of the equation of state and the extra-mixing. We considered both instantaneous mixing in the form of overshooting, using either the adiabatic or radiative temperature gradient as well as turbulent diffusion using two sets of values for the parameters of the turbulent diffusion coefficient. It appears that the overall best agreement was obtained for a model including values from \citet{Proffitt} for the parametric form of turbulent diffusion. However, the remaining discrepancies in the inversions seem to indicate that besides chemical mixing, a proper reproduction of the transition of the temperature gradients at the BCZ is also required. This confirms that more realistic implementation of overshooting may also influence the solar modelling problem. Moreover, the low position of the BCZ in these models may suggest that the modification of the opacity around the iron peak could be slightly too high if the updated Neon abundances are used.

Overall, our tests suggest that the solar problem does not originate from one single source, but rather various small contributors. First, additional investigations should be done to improve the formalism and hypotheses of microscopic diffusion \citep[see e.g.][]{Turcotte,Schlattl, Gorshkov08, Gorshkov10}. Comparisons between evolutionary codes from various groups \citep[such as in][]{ESTA1, ESTA2} could be pushed at the level of accuracy of helioseismology and provide crucial insights by pinpointing the intrinsic differences between each model, regardless of their (dis)agreement with the Sun. Second, further analyses using higher degree modes such as those derived from recent MDI datasets \citep{Reiter2015} could prove very useful to revisit studies such as those of \citet{DiMauro2002, Lin05, Vorontsov13} to constrain the upper parts of the convective envelope, the equation of state and the abundance of heavy elements \citep[see also][for recent seismic determinations]{VorontsovSolarEnv2014, BuldgenZ}, which are key elements of the issue.

Moreover, a strong limitation of our study is the absence of a treatment of the transition in temperature gradient in the BCZ. Combining the inversions presented here to analyses as in \citet{RoxVor94PhaseShift, JCDOV} could add strong constraints on the opacity changes just below the envelope. Indeed, coupling both techniques could probably help disentangling between the contributions of the opacity underestimation and the inadequate modelling of the temperature gradient transition in the tachocline to the overall discrepancies seen in the inverted profiles.

The recent investigations by \citet{JCD18} indicate that a density dependent diffusion coefficient coupled with significant modifications of the opacity profile (through both rescaling the metal mixture but also modifying the mean Rosseland opacity) can be used to eliminate the glitch of the tachocline in the solar data. From our tests, it appears that our formulation for the additional diffusive mixing coupled with our modification to the opacity profile does not erase completely the signal of the tachocline and is insufficient to reconcile the entropy plateau in the solar model with that of the Sun.

This could indicate that some entropy mixing could be present below the convective zone. However, this is purely speculative and further tests and comparisons should first be performed on the dependency of the form and intensity of the additional mixing introduced on the formalism, as well as on the hypotheses and consistency of the mixing of chemical elements already present in the standard solar models.

In that sense, the study of \citet{JCD18} complements well our work. Here, we have been able to show that inputs from additional inversions can provide further constraints on the solar problem. However, we stress again that such studies should be complemented by in-depth investigations of the reliability of solar models as in \citet{Boothroyd03} for various modelling groups.

With the recent g-modes detection by \citet{Fossat}, further studies could try using these constraints along with the frequency ratios of \citet{RoxburghRatios} and neutrino fluxes. However, such analyses should perhaps first remain unconnected, as the g-mode detection seems to be quite fragile \citep{Schunker2018} and trying to combine all constraints might be very challenging.

\section*{Acknowledgements}

AM and GB acknowledge support from the ERC Consolidator Grant funding scheme ({\em project ASTEROCHRONOMETRY}, G.A. n. 772293). We gratefully acknowledge the support of the UK Science and Technology Facilities Council (STFC). S.J.A.J.S. is funded by the Wallonia-Brussels Federation ARC grant for Concerted Research Actions. The work of V.K.Gryaznov, I.L.Iosilevskiy and A.N.Starostin was supported by RAS Scientific Program I.13 "Condensed matter and plasma under high energy density conditions". This article used an adapted version of InversionKit, a software developed within the HELAS and SPACEINN networks, funded by the European Commissions's Sixth and Seventh Framework Programmes. PE and GM acknowledge support from the Swiss National Science Foundation through the grant 200020-172505. The work of OR was supported by the "Programme National de Physique Stellaire" (PNPS) of CNRS/INSU co-funded by CEA and CNES. We wish to thank W. Däppen for fruitful discussions on questions related to the theory of equations of state used in stellar structure and evolution.

\bibliography{biblioarticleSun}

\end{document}